\documentclass{article}

\usepackage{arxiv}

\usepackage[utf8]{inputenc} 
\usepackage[T1]{fontenc}    
\usepackage{hyperref}       
\usepackage{url}            
\usepackage{booktabs}       
\usepackage{amsfonts}       
\usepackage{nicefrac}       
\usepackage{microtype}      
\usepackage{lipsum}		
\usepackage{graphicx}
\usepackage{natbib}
\usepackage{doi}

\title{Collective Behavior of Clusters of Free-to-Move Cylinders in the Wake of a Fixed Cylinder} 

\author{ \hspace{1mm}Daniela ~Caraeni\\ \\
	Department of Mechanical and Industrial Engineering\\
	University of Massachusetts Amherst \\
	Amherst, Massachusetts \\
	\texttt{dcaraeni@umass.edu} \\
	\And
	\href{https://orcid.org/0000-0002-7890-1699}{\includegraphics[scale=0.06]{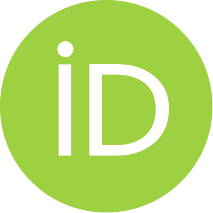}\hspace{1mm}Yahya ~Modarres-Sadeghi} \thanks{Corresponding author: modarres@engin.umass.edu)} \\ \\
	Department of Mechanical and Industrial Engineering\\
	University of Massachusetts Amherst \\
	Amherst, Massachusetts \\
	\texttt{modarres@engin.umass.edu} \\
}

\begin{document}

\maketitle

\begin{abstract} We study the collective behavior of clusters of cylinders placed in the wake of a fixed cylinder and free to move in a direction perpendicular to that of the incoming flow, with no structural damping or stiffness. We keep the Reynolds number, defined based on the cylinder diameter, at 100 and consider five different configurations for the initial positions of the cluster cylinders: linear, rectangular, V-shaped, triangular, and circular. In each configuration, we consider progressively increasing number of cylinders in the cluster. We show that overall, the cylinders tend to form final linear configurations, in which, after their transition, the cylinders form one or more lines. Some free-to-move cylinders might take the lead position in some of these linear formations depending on the initial configuration. These steady-state positions are achieved when the mean value of lift that acts on the cylinders becomes negligible. As a byproduct of these reconfigurations, the overall drag force that acts on the collection of cylinders reduces at their final steady-state locations in comparison with their original configurations. The complicated wakes that are observed in the fixed counterparts of these configurations are replaced by a series of vortex rows in the wake of separate lines of cylinders. Reducing the mass ratio allows the cylinders to oscillate about their mean displacement paths, but their transient paths and their final steady-state positions are not affected significantly by the decrease in the mass ratio.

\end{abstract}

\section{Introduction}

We consider the collective behavior of a group of bluff bodies with circular cross-sections (cylinders) placed in the wake of a fixed cylinder. When a bluff body with a circular cross-section is placed in flow, vortices are shed in its wake. These vortices exert fluctuating flow forces on the structure itself, and any other structure that is placed in close vicinity of the original bluff body. The interactions between the vortices that are shed from one bluff body and structures placed in its vicinity have been studied extensively in the past. Typical examples of such studies are cases in which flow behavior around two (or more) cylinders placed in tandem, side-by-side, or in a staggered configuration are studied, where all cylinders are fixed and the focus is mainly on the flow behavior around them. There are also cases where some or all of the cylinders placed in tandem or side-by-side arrangement are flexibly mounted and therefore can oscillate due to their interactions with the fluctuating forces caused by the vortices that are shed from the bluff body itself or the vortices shed from the neighboring bluff bodies. In this work, we consider cases where the cylinders that are placed in the wake are free to move in a direction perpendicular to the direction of the incoming flow, without any spring or damper attached to them. These cylinders are only influenced by the flow forces.

When rigid cylinders are placed in tandem, the gap between the cylinders is the main contributor to changes in the flow behavior around the cylinders~\citep{Carmo2011, Dehkordi2011}. For $1<L/D<1.5$ (where $D$ is the cylinder diameter and $L$ is the center-to-center distance between two cylinders), the shear layer from the upstream cylinder goes over the downstream cylinder and the flow in between the two cylinders stays almost still. Then, for $1.5<L/D<4$, the shear layers from the upstream cylinder reattach on the downstream cylinder, and for $L/D>4.5$, both cylinders shed vortices in their respective wakes~\citep{zdravkovich1977review,Igarashi1981,Igarashi1984,alam2008strouhal,alam2018vortex}. For smaller gap sizes (smaller than $3D$), a negative drag force acts on the downstream cylinders~\citep{Meneghini2001}. In side-by-side arrangements also, the gap size is the main parameter that influences the flow behavior~\citep{Hesam2011}. For small $H/D$ (where $H$ is the vertical distance between centers of two adjacent cylinders), a single bluff-body vortex shedding is observed---the two cylinders act as one rigid body. At medium $H/D$ values, biased flow with synchronized vortex shedding is observed, and at high $H/D$, symmetric flow with synchronized vortex shedding~\citep{sumner1999fluid}. For smaller gap sizes, i.e., $H/D \leq 2$, there is a repulsive force between the cylinders~\citep{Meneghini2001}. When several fixed cylinders are placed next to each other~\citep{Nicolle2011,Klettner2019}, if they are placed very close to each other, the collection behaves very similarly to a large bluff body. When the cylinders are placed very far from each other, they act as individual cylinders and shed vortices in their wakes. It is only the intermediate distances that cause interactions among wakes of different cylinders such that the details of the flow behavior in between the cylinders also influence the system's overall behavior.

When structures placed in the wake of a bluff body are allowed to oscillate, i.e., when they are flexibly mounted, wake-induced vibrations (WIV) are observed. The vortices that are shed from the upstream body interact with the bodies in the wake and exert external fluctuating forces on them, resulting in oscillations of structures in the wake of an upstream bluff body. Similar to the case of rigid bodies placed in the wake of an upstream body, it has been shown for WIV cases that the distances between the upstream and downstream structures and the relative locations of the bodies with respect to each other influence their responses~\citep{Fukushima2021,Hishikar2022, Meneghini2001, Mittal1997, Mittal2001, Prasanth2009, Skonecki2023, Kitagawa2008}.

The question that we ask in the present work, however, is not concerned with fixed or flexibly mounted bodies in the wake of an upstream body. The question is how will a group of cylinders that are free to move in the transverse direction (and are not attached to any spring or damper) behave in the wake of a bluff body? We keep the upstream cylinder rigid and fixed at all times. Naturally, a von Kármán vortex street is formed in the wake of this fixed cylinder. Then in this wake, we place several cylinders, initially located in different configurations, that are free to move in the transverse direction. The cylinders are not attached to any spring or any structural damper. They are completely free to move in the transverse direction, but they always stay at the same horizontal distance (in the direction of flow) from the cylinder. Had they not been limited in that direction, all the cylinders would have moved with the flow downstream. This configuration can be thought of as the bluff body equivalent of configurations studied in the cases of collective swimming or collective flying. In collective swimming or flying, the structures (the fish or the birds) are active structures that propel themselves forward by producing thrust forces, as they interact (sometimes beneficially and sometimes not) with the wake of their upstream fish or birds. In the system we consider here, bluff bodies are passive structures (they do not swim or fly), and they passively interact with the wake of their upstream structure. The passive nature of the cylinders in our case then requires them to be externally ``helped'' to stay in place in the direction of flow, which we do by restricting their motion to be in the transverse direction only.

The initial configurations that we have chosen for these cylinders are inspired by those used by the fish and birds in their collective swimming and flying~\citep{Bajec2009} but do not necessarily closely follow the configurations the active animals utilize. The parameter space that we explore is very large. The distances between the cylinders, the exact initial configurations, the number of cylinders in each case, the cylinder's mass ratio, and the Reynolds number are only some of the parameters that one can change in such a system. While in this work we consider many sets of parameters, our goal is not to explore the entire parameter space---that would have been an impossible task---but rather to give a view of the collective behavior of bluff bodies in the wake of a fixed bluff body.
 
\section{Problem formulation}

\subsection{Governing equations and numerical methods}
We consider the two-dimensional, incompressible cross-flow around multiple cylinders. The fluid flow is governed by the unsteady, incompressible Navier-Stokes (N-S) equations:
\begin{equation}
    \underline{\nabla} \cdot \underline{u}=0,
\end{equation}
\begin{equation}
    \rho(\frac{\partial\underline{u}}{\partial t} + \underline{u}\cdot\underline{\nabla} \, \underline{u})=-\underline{\nabla}\, p + \underline{\nabla}\cdot\underline{\underline{\tau}}.
\end{equation}

The finite volume method is used to discretize the unsteady N-S equations which are then solved using a high-resolution advection scheme and a second-order backward Euler transient scheme. The high-resolution advection scheme was chosen because it uses a second-order scheme when possible and blends into a first-order scheme only to remain bounded. This scheme gives higher accuracy because high-resolution advection schemes result in less numerical diffusion and less artificial damping of the solution. This, coupled with selecting the second-order backward Euler scheme, keeps the solution close to second-order accuracy. The convergence tolerances for the continuity and velocity components are set to an RMS value less than $10^{-4}$. 

\subsection{Domain setup}

\begin{figure}
    \centering
    \includegraphics[width=\textwidth]
    {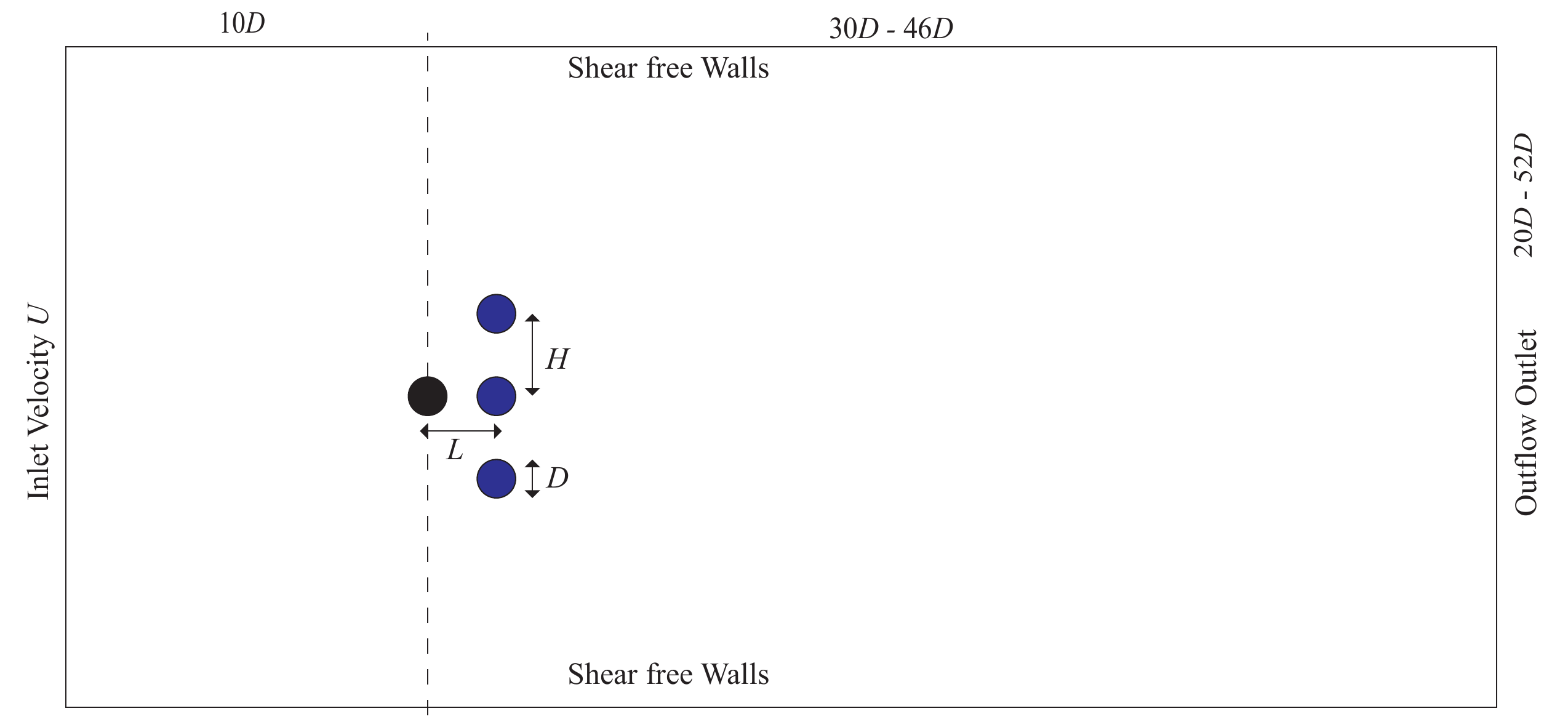}
    \caption{A schematic of the setup used here. The fixed cylinder is shown in black, and the cylinders that are free to move in the vertical direction are shown in blue. The distances between consecutive cylinders are $L$ in the horizontal direction and $H$ in the vertical direction.}
    \label{fig:domain_BCs}
\end{figure}

Figure~\ref{fig:domain_BCs} illustrates a general layout of a lead cylinder with follower cylinders placed at distances $L$ and $H$ away from the lead cylinder. This configuration is placed in a 40D$\times$20D domain that is meshed with an unstructured grid with a total of over 75000 grid elements. 
A uniform flow is introduced at the inlet with an outflow outlet on the opposite side. A shear-free boundary condition is applied at the top and bottom walls. At the cylinder wall, a no-slip condition is applied. The cylinder diameter is $D$ = 1 cm, and the mass ratio, defined as the mass of the cylinder over the mass of the displaced fluid is kept constant at $m^*=12.7$ for the first series of the results (the ``high'' mass ratio cases) and then at $m^*=1$ for the ``low'' mass ratio cases. The cylinder is placed $10D$ downstream of the inlet and $20D$ to $52D$ from the upper and lower walls, depending on the case. The Reynolds number defined based on the diameter of the fixed cylinder, which is identical to the diameters of all the freely moving cylinders, is kept constant at  $Re=100$. The time step is set to 0.001 s.

We use dynamic mesh with smoothing and remeshing methods due to the mesh motion that would be associated with the large displacements of the free-to-move cylinders and their oscillations. For smoothing, we use the diffusion method to keep the mesh quality as the cylinders relocate in the domain. For remeshing, we use methods-based remeshing, which allows us to select the minimum and maximum length scales for remeshing as well as a maximum cell skewness and a maximum face skewness which we keep to 0.55 and 0.7, respectively. Each configuration follows the same meshing strategy but the domain is increased as the number of cylinders is increased, resulting in up to 150000 nodes and 300000 elements in the mesh. 

\begin{figure}
    \centering
    \includegraphics[width=\textwidth]
    {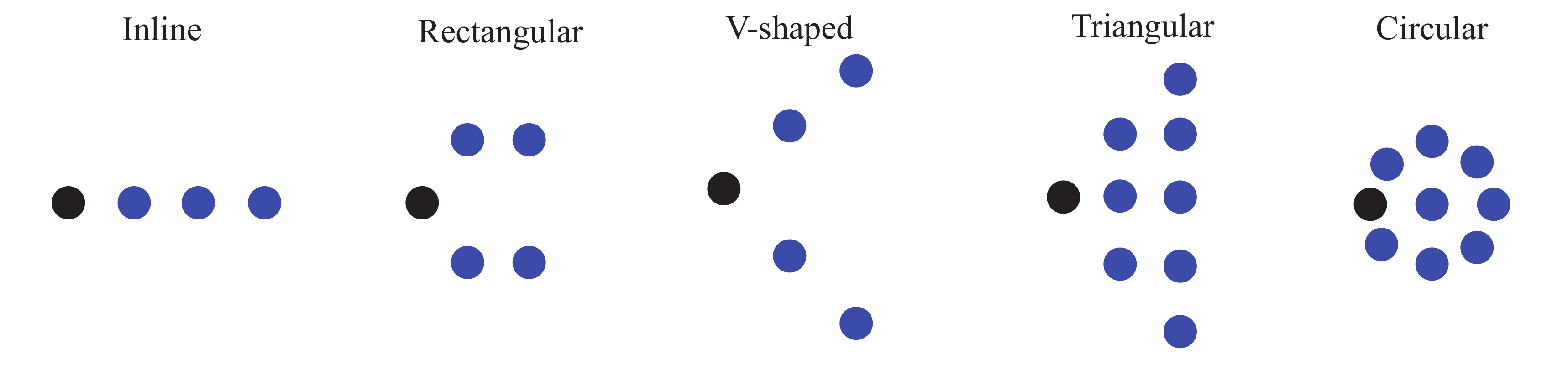}
    \caption{Five different configurations that we consider for the clusters of cylinders in this study.}
    \label{fig:new_diff_configurations}
\end{figure}

We show a schematic of the initial configurations that we use in the study in Figure~\ref{fig:new_diff_configurations}: The ``inline'', the ``rectangular'', the ``V-shaped'', the ``triangular'', and the ``circular'' configurations. Note that the free-to-move cylinders take the form of these configurations at time zero when there is no flow. As soon as the flow starts, the transient motion of the cylinders starts. This motion is what we will consider in this work.

\subsection{Verification}
\begin{figure}
    \centering
     \includegraphics[width=0.5\textwidth]
     {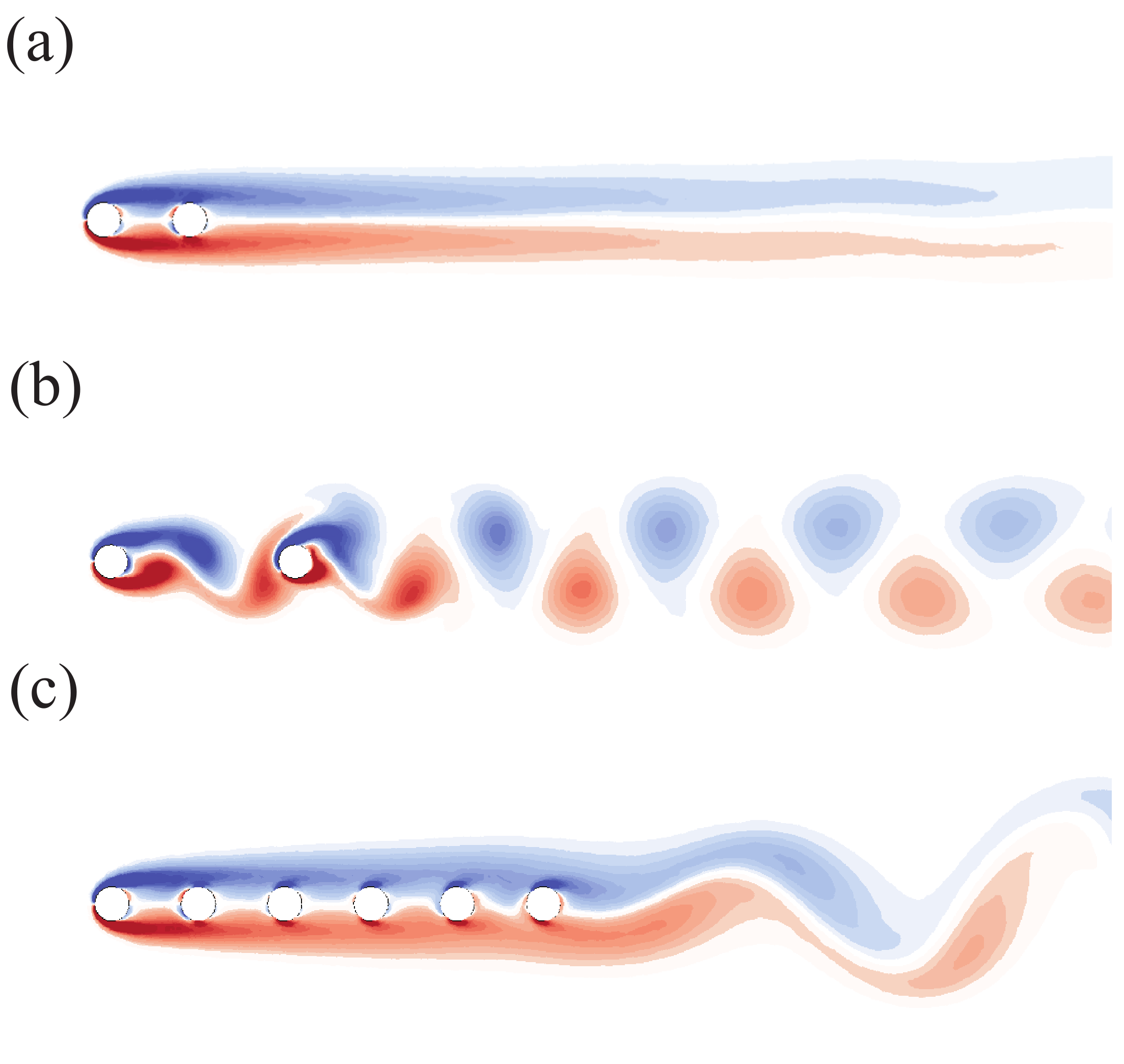}
    \caption{Vorticity contours for validation cases with (a) two fixed cylinders in tandem with $L/D = 2.5$, (b) and $L/D = 5.5$, same as cases considered by \cite{Ding2007}, and (c) six fixed cylinders in tandem with $L/D = 2.5$, same as the case studied by \cite{Hosseini2020}.}
    \label{fig:validation_vorticity_contours}
\end{figure}

We first ran a series of cases for fixed cylinders and compared our results with previously published results. For these runs, we considered the case of a single cylinder, two cylinders placed in tandem with $L/D=2.5$ and 5.5, and six cylinders placed in tandem with $L/D=2.5$, all at $Re=100$. A summary of these comparisons is given in Tables~\ref{tab:2Cyl_L_D_2_5_and_5_Lit} and \ref{tab:6Cyl_L_D_2_5_Lit} in the form of the mean drag coefficient and the fluctuating lift coefficient, and a general agreement between our results and these previous results is observed. In addition to the flow forces that act on each cylinder, we qualitatively compared the wake that is formed behind these sample cases (shown in Figure~\ref{fig:validation_vorticity_contours}) with those observed in the literature. In the case of two tandem cylinders placed at $L/D=2.5$, no shedding is observed between the two cylinders, while when the two cylinders are placed at a distance of $L/D=5.5$, vortices are shed both in between the two cylinders and in the near wake of the second cylinder. This is in agreement with the observation by \cite{Kitagawa2008} and \cite{Ding2007}. In the case of the six cylinders placed in tandem, large vortices form in the wake, similar to what \cite{Hosseini2020} have observed.

\begin{table}
    \centering
    \begin{tabular}{ccccc}
      Single cylinder &  & Current Study  & \cite{Calhoun2002} & \cite{WangFan2009} \\
    \hline
               & $C_{D,mean}$ & 1.35 & 1.33  & 1.379\\
        & $C_{L,fluct}$ & 0.31 & 0.298 &  0.357 \\
        
    \hline
     Two cylinders in tandem & & Current Study & \multicolumn{2}{c} {\cite{Ding2007}} \\
    \hline
    $L/D$ = 2.5; Cylinder 1 &
        $C_{D,mean}$ & 1.19 &\multicolumn{2}{c} {1.163}\\
         & $C_{L,fluct}$ & 0.0016 &\multicolumn{2}{c} {0.00} \\
    $L/D$ = 2.5; Cylinder 2 &
        $C_{D,mean}$ & -0.05 &\multicolumn{2}{c} { -0.0895} \\
         & $C_{L,fluct}$ & 0.0002 &\multicolumn{2}{c} {0.00} \\
    \hline
    $L/D$ = 5.5; Cylinder 1:  &
        $C_{D,mean}$ & 1.32 &\multicolumn{2}{c} {1.329}\\
        & $C_{L,fluct}$ & 0.33 &\multicolumn{2}{c} {0.330} \\
    $L/D$ = 5.5; Cylinder 2: &
        $C_{D,mean}$ & 0.9 &\multicolumn{2}{c} {0.858} \\
        & $C_{L,fluct}$ & 1.5 &\multicolumn{2}{c} {1.554} \\
    \end{tabular}
    \caption{The values of mean drag coefficients and fluctuating lift coefficients acting on the cylinders calculated in the current study in comparison with the previously published results.}
    \label{tab:2Cyl_L_D_2_5_and_5_Lit}
\end{table}

\begin{table}
    \centering
    \begin{tabular}{ccc}
    Comparison of $C_{L}$ & Current Study & \cite{Hosseini2020} \\
    \hline
    Cylinder Number: \\
        1 & 0.001 &  0.003 \\
        2 & 0.003 &  0.006  \\
        3 & 0.02  &  0.014  \\
        4 & 0.02  &  0.014  \\
        5 & 0.05  &  0.035  \\
        6 & 0.2   &  0.2
    \end{tabular}
    \caption{The values of fluctuating lift coefficient for cylinders 1 through 6 in tandem arrangement.}
    \label{tab:6Cyl_L_D_2_5_Lit}
\end{table}

\subsection{Normalized lift and drag coefficients}
To quantify the influence of reconfiguration on the flow forces that act on the cylinders, in this study we define the normalized lift and drag coefficients as 

\begin{equation}
    C^*_L = \frac{2 F_L}{N \rho U^2 A}, 
\end{equation}

\noindent
and

\begin{equation}
    C^*_D = \frac{2 F_D}{N \rho U^2 A},
\end{equation}
where $F_D$ and $F_L$ are the summations of all lift and drag forces acting on all cylinders in any given configuration, respectively, $N$ is the number of cylinders in the configuration (including the fixed cylinder), $\rho$ is the density of the fluid, $U$ is the flow velocity, and $A$ is the characteristic area. For each case where we conduct this study of flow forces, we run a case of cylinders fixed at their initial configuration as well. We calculate the normalized lift and drag coefficients for the rigid case as well as the case where all cylinders except the very first one are free to move, and then compare the values to quantify how the reconfiguration of cylinders has influenced the overall flow forces that the system experiences.

\section{Inline Configurations}

\begin{figure}
    \centering
    \includegraphics[width=0.8\textwidth]
    {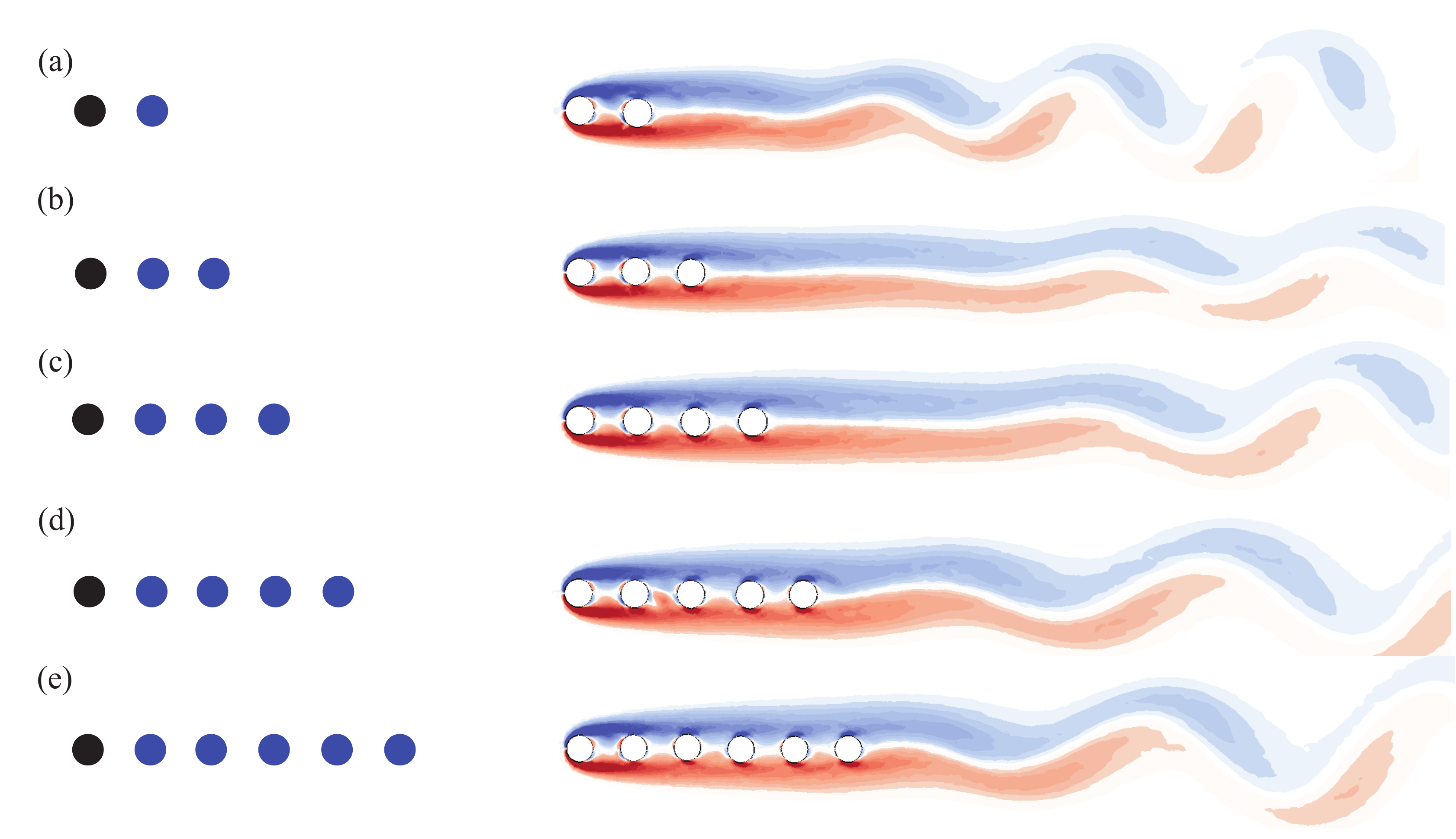}
    \caption{The ``Inline Configurations'' for an increasing number of cylinders and the flow behavior in their wakes. The black cylinder is fixed and the blue cylinders are free to move in the vertical direction.}
    \label{fig:Inline_configurations}
\end{figure}

We start by examining the inline configuration, in which the initial locations of all cylinders are behind the fixed cylinder (Figure~\ref{fig:Inline_configurations}). For all these cases, the front cylinder is fixed and disturbs the incoming flow. The cylinders that are placed in the wake of this cylinder are free to move in the direction perpendicular to the direction of flow, i.e., the $y$-direction. We start by placing only one cylinder in the wake of the rigid cylinder and we increase the number of cylinders in the wake to 5. The cylinders are kept at $L/D=2$, for consistency, and are numbered 1 through 6 from left to right, when 1 corresponds to the fixed cylinder.

When only one cylinder is in the wake of the fixed cylinder, it does not oscillate much (its maximum amplitude of oscillations is $A^*=A/D=0.009$). For this $L/D$, the follower cylinder is sufficiently close to the fixed cylinder that the shear layers that leave the fixed cylinder do not form a vortex before reaching the second cylinder, and as a result, the second cylinder interacts with the shear layers that have left the fixed cylinder. This leads to the formation of long shear layers that cover both cylinders and vortices that are shed far from the second cylinder. The wake of the two cylinders seems to have merged, causing the behavior of the wake to resemble that of an equivalently longer body. Previous studies have shown that when two cylinders are placed in tandem, the formation of steady wake flow is observed for $L/D \leq 2$ at a Reynolds number $Re < 100$ (\cite{Singa2010, Dehkordi2011}). It was shown that the flow only becomes unsteady within the gap when $L/D \geq 3$. At $Re = 100$, no distinct vortex shedding has been observed behind the upstream cylinder (i.e., in the gap between the two cylinders). Instead, the shear layers reattach to the downstream cylinder, inhibiting the vortex shedding within the gap~\citep{Mittal1997}, similar to what we observe here. 

As we add more free-to-move cylinders to the wake of the fixed cylinder, a similar scenario is observed. In all cases, the shear layers that leave the fixed cylinder encompass the cylinders in the wake, and vortices are shed in the wake of the last cylinder. Quantitative differences, however, are observed in the behavior of cases with more cylinders in the wake. The amplitude of oscillations of the free-to-move cylinders starts increasing as the number of cylinders is increased. The maximum amplitude of oscillations of a cylinder in the inline configuration changes from $A^*=0.04$ to $A^*=0.13$, $A^*=0.09$, and lastly $A^*=0.21$, respectively for 2, 3, 4, and 5 cylinders in the wake. While the amplitudes for the first three cases are relatively small (around $0.1D$), the amplitude becomes much larger in the last case (around $0.2D$). In cases where oscillations are observed in the wake, they are observed to start from the last cylinder in the row and influence the cylinders upstream. In the inline configuration, although the magnitude of oscillations of the follower cylinders increases as the number of cylinders is increased, the cylinders keep their inline configuration, and none of them experiences enough oscillations to break out of its position.

\section{Rectangular Configurations}

\begin{figure}
    \centering
    \includegraphics[width=\textwidth]
    {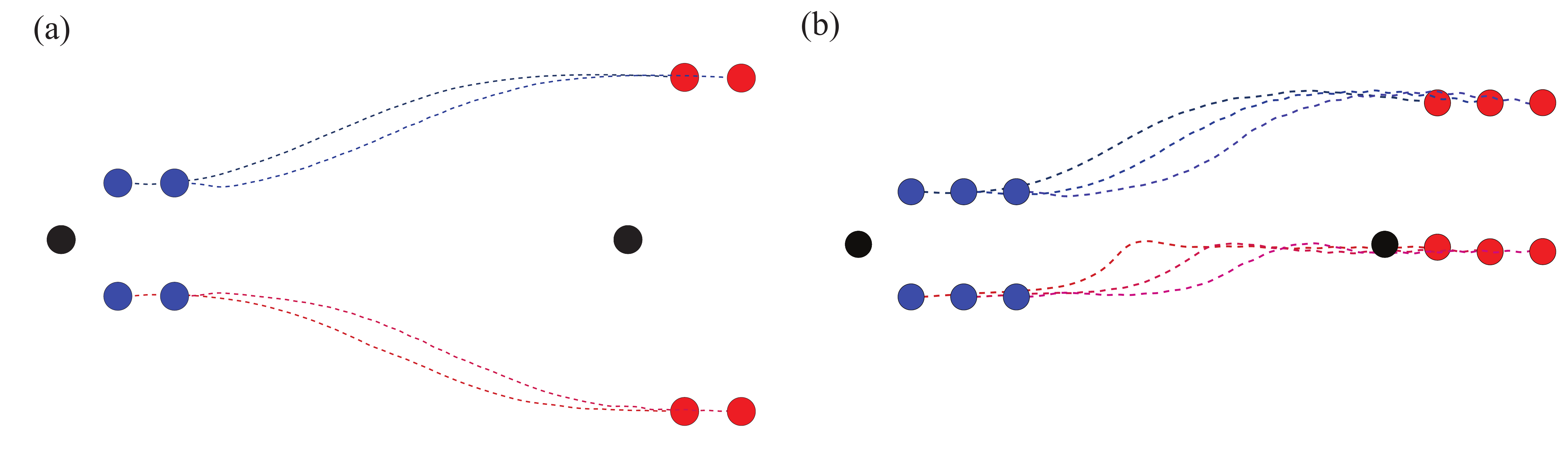}
    \caption{The initial (blue) and the final (red) locations of two cases of ``rectangular configurations'', in which two series of inline cylinders (two cylinders in each row in one case and three cylinders in each row in the other case) are located in the wake of a fixed cylinder, at equal vertical distances from the rigid cylinder. The dashed lines show the paths that the cylinders take from their initial locations to their final locations.}
    \label{fig:rectangle_configurations}
\end{figure}

In the linear configuration, the cylinders that were placed in the wake of the fixed cylinder could stay in between the shear layers that left the fixed cylinder and did not interact with the vortices that would have been shed from the wake of a rigid cylinder. Here we consider a configuration that includes cylinders that are placed inline, similar to the previous case, but they form two parallel lines and are initially located at constant vertical distances from the fixed cylinder (Figure~\ref{fig:rectangle_configurations}). We refer to this configuration as the ``rectangular configuration", and consider two cases for it: One in which each row of the linear cylinders consists of 2 cylinders and one in which there are 3 cylinders in each row. In this configuration, the free-to-move cylinders are located at a vertical distance of $H/D=2$ from the fixed cylinder, and their horizontal distances are kept at $L/D = 2$.

\subsection{The overall transient behavior}

The initial locations of the cylinders, the paths they take as they interact with the incoming flow, and their final locations in these two cases for this configuration are shown in Figure~\ref{fig:rectangle_configurations}. In both cases, the cylinders that start inline, stay inline. In the case with two cylinders in each row (Figure~\ref{fig:rectangle_configurations}(a)), the cylinders move away from their initial locations with respect to the rigid cylinder, i.e., $H/D=2$. The upper cylinders settle at a new location of $H/D=5$ and the lower cylinder at a location of $H/D=-6$. Note that while the initial locations of the two rows of the cylinders were at the same vertical distances from the fixed cylinder, their final locations were not. In the case with 3 cylinders in each row (Figure~\ref{fig:rectangle_configurations}(b)), the upper cylinders move outward and settle at $H/D=5$, and the lower cylinders move inward and reach $H/D=0$, i.e., they settle right behind the fixed cylinder and resemble a case of inline cylinders as discussed in the previous section. The behaviors of these two cases raise several questions, including two major ones: (i) Why is it that while the system is symmetric and the initial vertical distances of the two rows of cylinders from the fixed cylinder are equal to each other, the final configuration in none of the cases is symmetric? (ii) Why is it that in the case of two cylinders in each row, the cylinders in both rows move away from the rigid cylinder, but in the case of three cylinders in each row, one row moves toward the center?

A major difference between the rectangular configuration and the inline configuration is that in the inline configuration, there is only one ``leader'' cylinder, and that is the fixed cylinder. The other cylinders stay in the wake of that single leader and do not leave their original location. In the rectangular configuration, however, there are three ``leader'' cylinders: one is the fixed cylinder and the other two are the cylinders to the extreme left of each row. The local behavior for each row of cylinders is very similar to that observed in the inline configurations discussed in the previous section. All cylinders in the same row stay within the shear layers that leave the ``leader'' of that row. The cylinders oscillate slightly, but they never leave the inline configuration. The global behavior of these two lines of cylinders, however, is influenced heavily by the vortices that are shed from the fixed cylinder. When the vortices are formed in the wake of the fixed cylinder, they create an asymmetric pressure distribution around the cylinders placed on both sides, which results in a force on the free-to-move cylinders in the vertical direction.

\begin{figure}
    \centering
    \includegraphics[width=\textwidth]
    {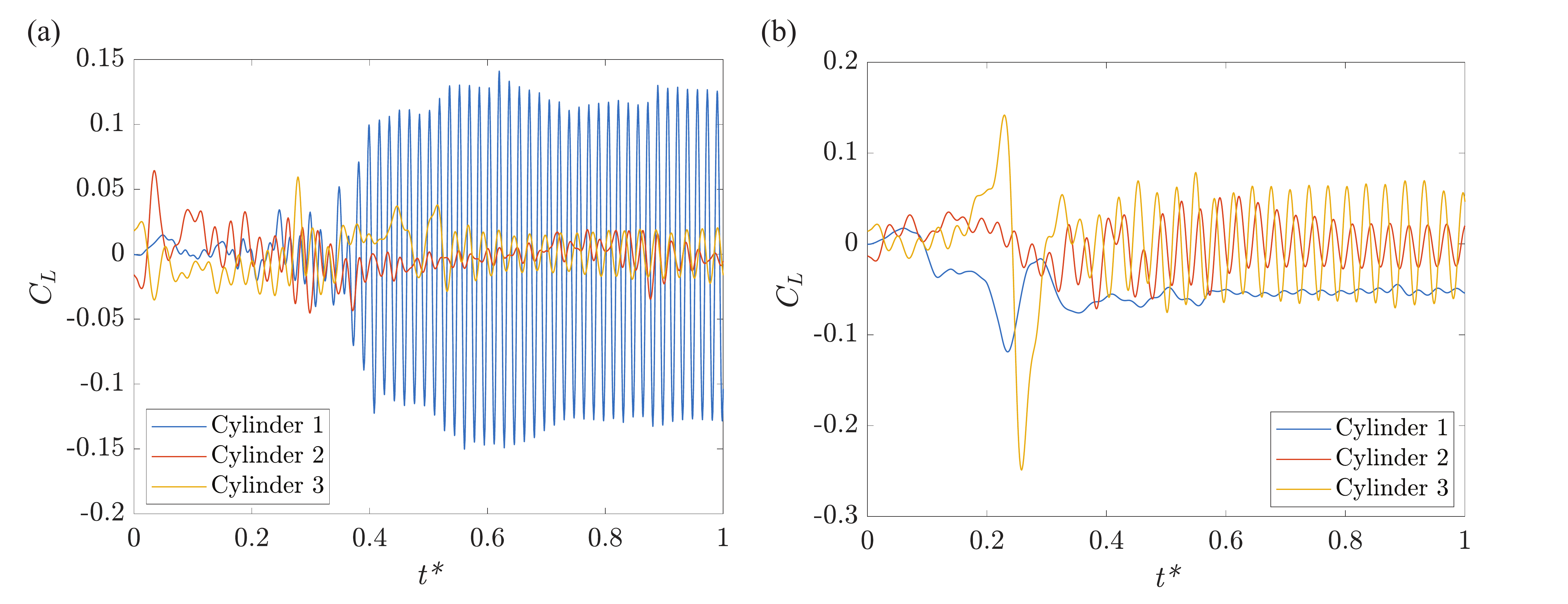}
    \caption{Lift coefficients for the first three cylinders in case (a) with a total of 5 cylinders and case (b) with a total of 7 cylinders. Cylinder 1 corresponds to the fixed cylinder, cylinder 2 is the first cylinder on the top row, and cylinder 3 is the first cylinder on the bottom row. In the plots, $t^*$ is the normalized time.}
   \label{fig:rectangle_configurations_lift_coefficient}
\end{figure}

\subsection{The flow behavior}

The transient forces that act on the cylinders from the time the vortices start to form in the wake of the fixed cylinder to the time that a steady-state shedding is formed in the wake of the cylinder cause the transition of the cylinders from their initial locations to their final locations. Since vortices are shed asymmetrically in the wake of the fixed cylinder, these transient forces that act on the free-to-move cylinders are not symmetric, which results in asymmetric final locations of the row of cylinders to the rigid cylinder. These forces are shown in Figure~\ref{fig:rectangle_configurations_lift_coefficient}. The asymmetry of forces that act on the two lead cylinders in the two rows is clear from these plots (the orange and yellow lines in the plots). It is this asymmetry in forces that results in asymmetric final locations of the cylinders. As the cylinders approach their steady-state conditions, the mean value of lift forces that act on the free-to-move cylinders becomes zero. As a result of this zero mean lift force, the cylinders settle in their steady-state conditions. The mean lift force on the fixed cylinder, however, is not necessarily zero. This is observed very clearly in case (b) of Figure~\ref{fig:rectangle_configurations_lift_coefficient} where the mean value of the lift force is a non-zero negative value while the free-to-move cylinders experience only fluctuating forces. In this case, had we let the fixed cylinder free in the transverse direction, it would have moved toward a steady state location where the mean lift would have been zero, and with itself, it would have moved the follower cylinders. We, however, do not let any of the lead cylinders free in this study!

\begin{figure}
    \centering
    \includegraphics[width=\textwidth]
    {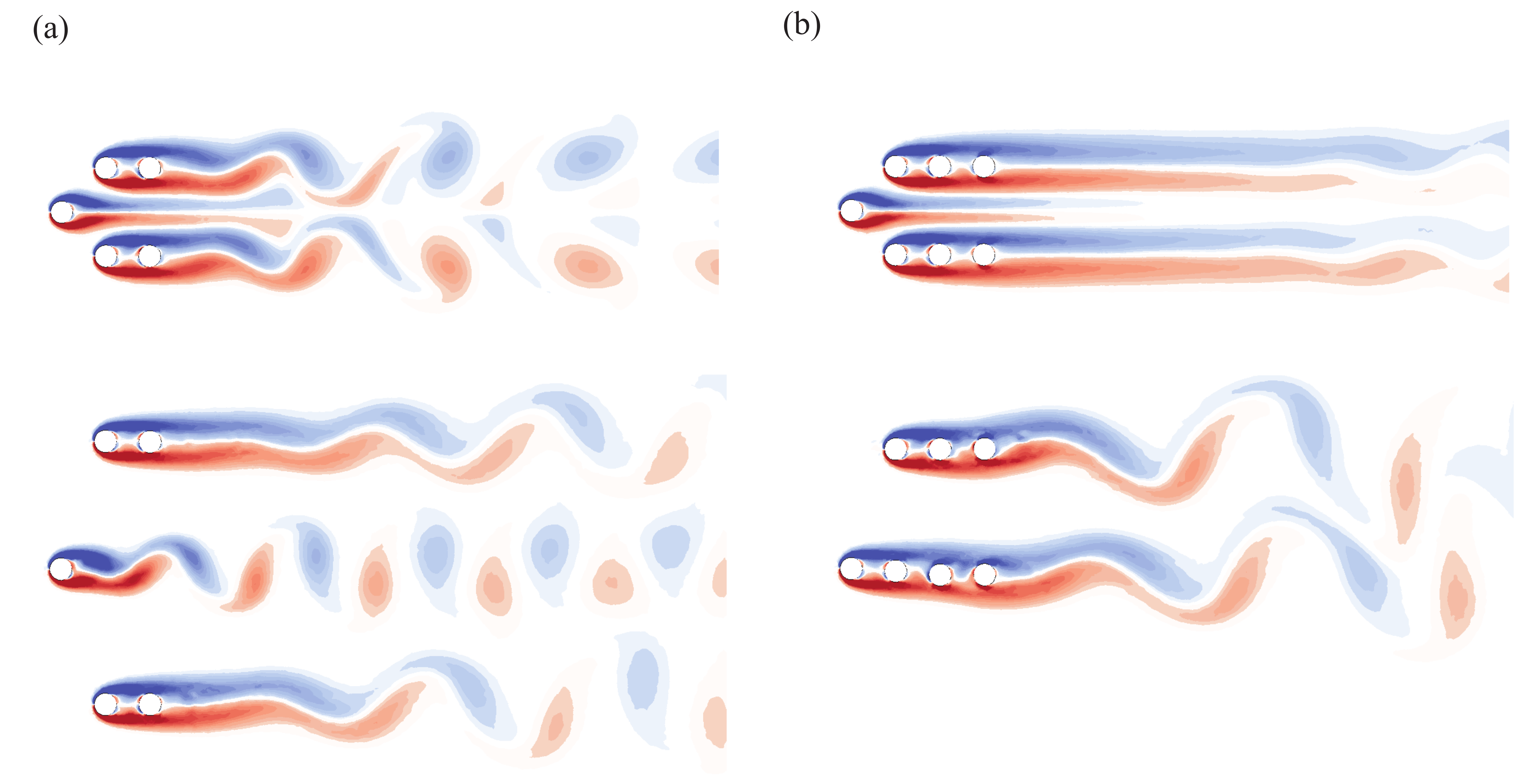}
    \caption{Vorticity contours for the two rectangular configurations shown in Figure~\ref{fig:rectangle_configurations}, for fixed configurations (upper row) and the final locations of the free-to-move cylinders (lower row).}
    \label{fig:rectangle_configurations_vorticity}
\end{figure}

The flow behavior around the cylinders in the two rectangular configurations is shown in Figure~\ref{fig:rectangle_configurations_vorticity} both for the cases where the cylinders are fixed and when they are free to move. When all cylinders are fixed, vortices are shed in the wake of the two rows of cylinders. The shear layers behind the fixed cylinder remain stable and do not form vortices. By adding a cylinder to the rows of cylinders, the shear layers in the wake of these rows are elongated, and the vortices are shed farther from the cylinders. When the cylinders are free to move, three rows of vortices are observed in the wake in Figure~\ref{fig:rectangle_configurations_vorticity}(a) and two rows in Figure~\ref{fig:rectangle_configurations_vorticity}(b). In the case with three rows of vortices in the wake, the shedding frequency for the middle vortices is higher than that for the other two, since the middle vortices are shed in the wake of a single cylinder, and the other two are shed in the wake of two cylinders. \cite{Xu2004} observed a drop in the Strouhal number as $L/D$ is increased from 1 to approximately 3, with the Strouhal number again increasing before plateauing at $L/D$ ranging from 7 to 15. In the case with two rows of vortices, the shedding frequencies are again different, because they are in the wakes of respectively 3 and 4 cylinders.

\subsection{The Importance of Initial Conditions}

\begin{figure}
    \centering
    \includegraphics[width=\textwidth]
    {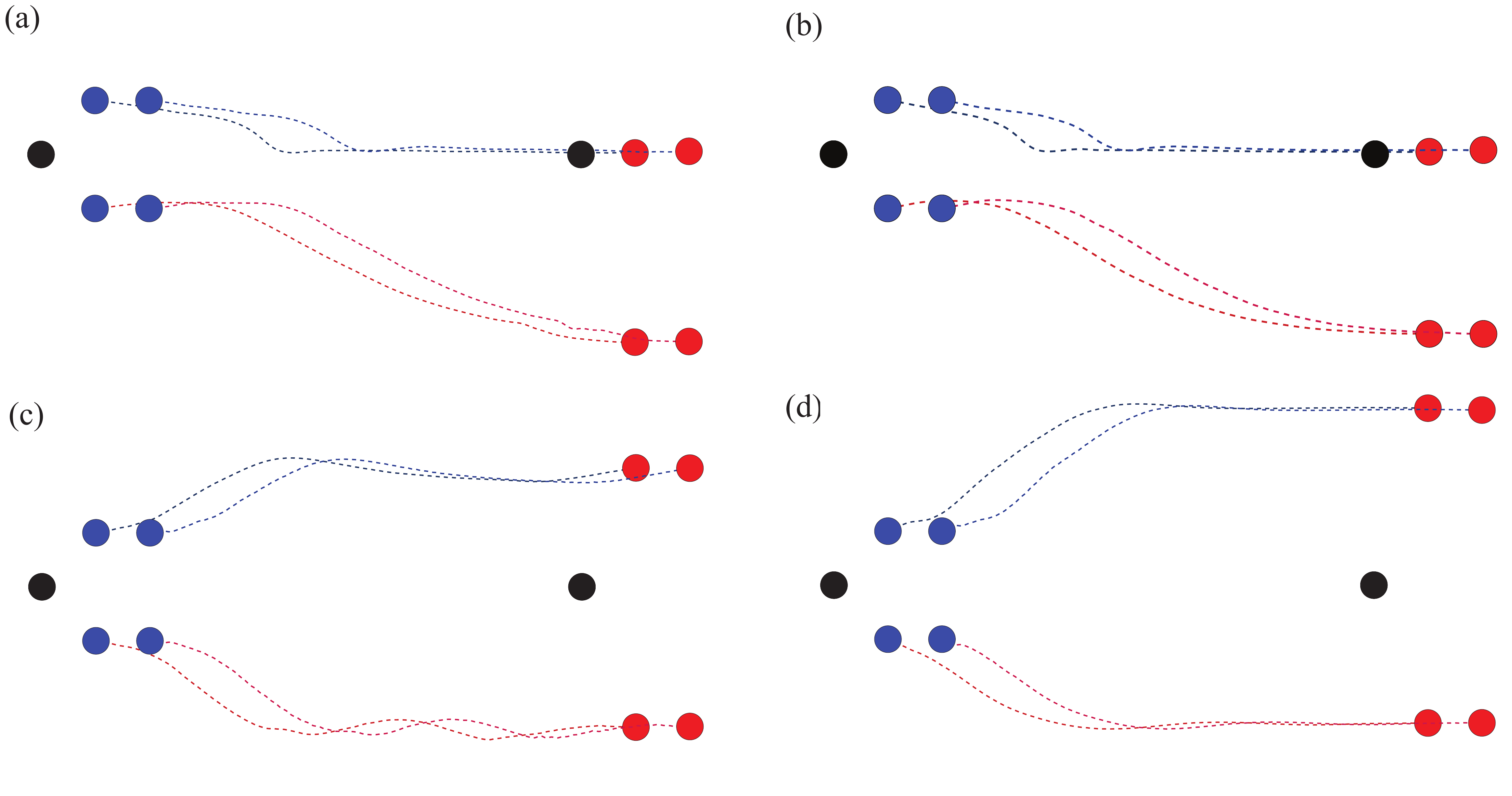}
    \caption{Rectangle configurations with externally imposed initial conditions, where both rows of cylinders are given an initial velocity inward: (a) $\dot{y} = 0.1D$/s and (b)  $\dot{y} = 0.2D$/s, and where both rows of cylinders are given an initial velocity outward: (c) $\dot{y} = 0.1D$/s and (d) $\dot{y} = 0.2D$/s.}
    \label{fig:rectangle_configurations_initialConditions}
\end{figure}

The behavior of cylinders in the rectangular configuration implies that the final locations of the cylinders in the wake depend on the initial conditions of the fluid around the lead cylinder in each row. This initial condition is the result of the details of the shedding of vortices in the wake of the fixed cylinder. These differences in the initial conditions could result in sending the cylinders away from the centerline, or toward the centerline, as we observed in the results we discussed in the previous section. In those cases, the initial condition is given passively and only as a result of the interactions between the shed vortices in the wake of the fixed cylinder and the free-to-move cylinders. Here, we control these initial conditions by providing initial velocities to the free-to-move cylinders to force a prescribed direction of motion. We do these tests for the case where two cylinders are placed in each row of the rectangular configuration. In the original, passive case both rows of cylinders moved outward (Figure~\ref{fig:rectangle_configurations}(a)).

Naturally, the first case that we test is a case where we give initial conditions to the cylinders that would move them toward the center of the wake---the opposite of what they did passively. This initial condition is imposed as an initial inward velocity of $\dot{y} = 0.1D$/s and then an initial inward velocity of $\dot{y} = 0.2D$/s that acts on each of the free-to-move cylinders. These values are comparable to the velocity that these cylinders experienced in the passive case that we discussed in the previous section. The responses of cylinders under these initial conditions are shown in Figure~\ref{fig:rectangle_configurations_initialConditions}(a) and (b). The upper row of cylinders follows the direction of the imposed initial conditions and moves inward. The lower row of cylinders, however, follows more or less the same path as the passive response and moves outward, despite the inward initial conditions. The upper row of cylinders moves to the center of the wake and makes up an inline arrangement behind the fixed cylinder. This results in a final configuration similar to what we had observed before in the case of three cylinders on each side. In that case, the flow (with no externally imposed initial condition on the cylinders) forced the lower row of cylinders into the wake of the fixed cylinder (Figure~\ref{fig:rectangle_configurations}(b)).

If the externally imposed initial velocity is given to the cylinders in an outward direction---the direction that the cylinders did move in the passive case, then it is not surprising to observe that the cylinders move in a very similar fashion to what they did in the passive case. If the initial velocity is comparable with the velocity that the cylinders experienced in their passive response (i.e., $\dot{y} = 0.1D$/s)), then their paths are very similar to their paths in the passive case (Figure~\ref{fig:rectangle_configurations_initialConditions}(c)). If, however, the externally imposed outward initial velocity is larger, i.e., $\dot{y} = 0.2D$/s then the cylinders again follow similar paths to the passive case, but move farther from the centerline of the wake, due to a larger initial velocity (Figure~\ref{fig:rectangle_configurations_initialConditions}(d)). The final locations of the rows of cylinders are not symmetric with respect to the centerline of the wake in any of these cases. For example, in Figure~\ref{fig:rectangle_configurations_initialConditions}(d), the upper row of cylinders is at a distance of $5D$ from the centerline, while the lower row is at a distance of $6.5D$. This asymmetric final location is due to the asymmetric forcing that acts on the lead cylinders during their transient motion---as we discussed in the previous section for the passive case.

\section{V-Shaped Configurations }

\begin{figure}
    \centering
    \includegraphics[width=\textwidth]
    {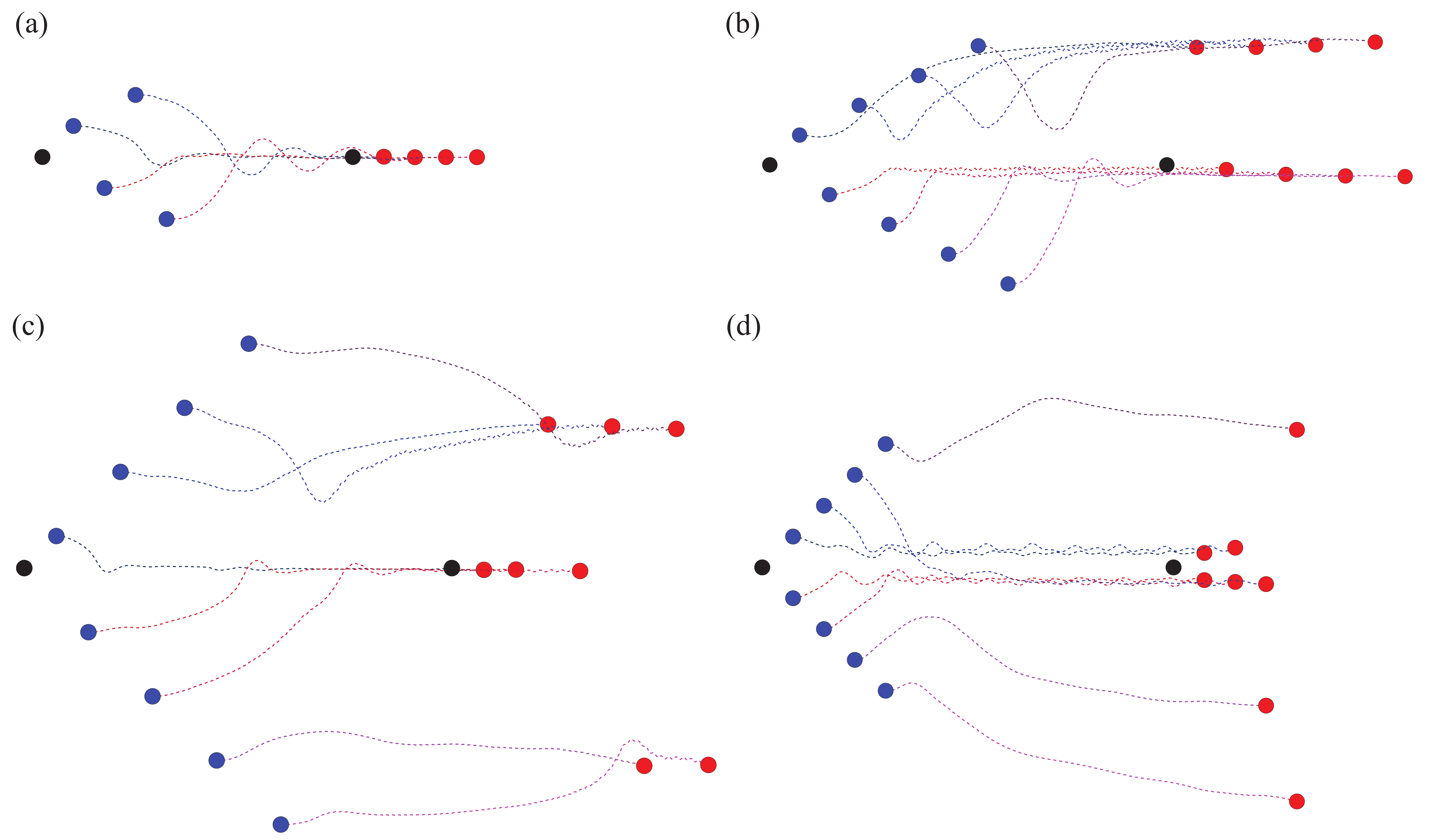}
    \caption{The initial (blue) and the final (red) locations of four cases of ``V-shaped configurations'' considered here. The dashed lines show the paths that the cylinders take from their initial locations to their final locations.}
    \label{fig:V-shape_configurations}
\end{figure}

In the past two configurations, inline and rectangular, the free-to-move cylinders were initially placed in one or two lines. The interactions of these cylinders locally with the shear layers of the lead cylinder resulted in them staying at the same inline configuration throughout their response. In the rectangular configuration, the cylinders moved from their initial locations to new locations, however, the cylinders in each row always stayed in line. The question then arises of what will happen if the free-to-move cylinders are not organized in a line initially. Will they stay in their initial locations relative to each other, will they form a line, or will they take any other final configuration? There are, clearly, several different ways that one can configure the initial locations of the free-to-move cylinders in the wake of the rigid cylinder such that they are not placed in a single line parallel to the direction of flow. Here, we start by looking at a case where the cylinders are located in the wake of the fixed cylinder in the form of a V---similar to the V-formation observed in bird flight. There are also several different ways this V-formation can be configured. Here, we consider four configurations. 

\subsection{V-shaped configuration --- Case (a)}

In configuration (a), we consider the case where the cylinders are kept at a horizontal distance of $L/D=2$ and a vertical distance of $H/D=2$ to each other. We then remove every other cylinder in each row, such that at every horizontal $n \times L/D$ location, there will only be one cylinder. We consider this configuration with two cylinders on each side as shown in Figure~\ref{fig:V-shape_configurations}(a). The reason we remove every other cylinder is to enable the cylinders to form a single line if that is the desired steady-state solution for the cylinders. As seen in the final state of this configuration in Figure~\ref{fig:V-shape_configurations}(a), this is indeed the desired response. All four cylinders move inward immediately after the transient response begins. The two cylinders closer to the fixed cylinder---and closer to the centerline---reach their steady-state condition behind the fixed cylinder directly, while the other two cylinders oscillate about the centerline before reaching their steady-state conditions. Eventually, all four cylinders form a line in the wake of the fixed cylinder. Note that in the initial condition for this form, all five cylinders could technically act as a lead cylinder, because none of them is completely placed behind another. However, as seen in the snapshots of Figure~\ref{fig:V-shape_configurations_fixed_moving}, the cylinders in the wake of the fixed cylinder are attracted to its wake and eventually the fixed cylinder becomes the sole lead cylinder. Once the inline configuration is formed, the cylinders stay in their locations and behave as they did in the inline configuration observed in Figure~\ref{fig:Inline_configurations}.

\subsection{V-shaped configuration --- Case (b)}

In this configuration, we increase the number of cylinders in the wake of the fixed cylinder to four cylinders on each side, following the same criteria for the distances between the cylinders, $L/D=2$ and $H/D=2$, and after removing every other cylinder on each side (Figure~\ref{fig:V-shape_configurations}(b)). The question is whether or not the formation of an inline configuration will be affected by the number of cylinders in the wake. As shown in the final configuration of cylinders, they again form inline configurations, however, this time there are two lines of cylinders---similar to one of the steady-state responses observed in the rectangular configuration. It is interesting that in this case, the closest free-to-move cylinder to the fixed cylinder moves outward initially, instead of inward. It is this movement of this cylinder that affects the movements of the cylinders that were originally located on that cylinder's side of the configuration, and they follow this cylinder as a lead cylinder and converge to their desired inline configuration behind this cylinder. The cylinders on the other side of the fixed cylinder move inward, and form a line behind the fixed cylinder. The final configuration then consists of two rows of cylinders, one behind the fixed cylinder and one behind a free-to-move cylinder.

\subsection{V-shaped configuration --- Case (c)}

Another question that arises is how much the behavior of cylinders in a V-formation depends on the initial distances between the free-to-move cylinders. If the cylinders are placed too far from each other, for example, at a distance of $20D$ in the horizontal and vertical directions, it is expected that the follower cylinders do not see any influence from the lead cylinders. But if they are placed relatively close to each other, will their initial distance influence the formation of the linear configuration at the end of the transient response? To answer this question, we modify the initial configuration of the V-formation such that the cylinders on the two sides of the fixed cylinder stay on a $45^{\circ}$ angle line. We place cylinders on such lines by placing them at distances of $L/D=2, 4, 6$ and $H/D=2, 4, 6$. We then remove every other cylinder on each side, for the same reason discussed before, to get the configuration shown in Figure~\ref{fig:V-shape_configurations}(c). Note that cylinders are placed farther from the centerline in this configuration in comparison with case (b). This longer distance then results in the fact that the cylinders that are closer to the fixed cylinder are still attracted to the centerline and form a line behind the fixed cylinder (with three free-to-move cylinders in the wake of the fixed cylinder), while the other cylinders that are farther from the centerline form independent linear formations on the two sides of the fixed cylinder. As a result, the end configuration consists of three linear configurations of cylinders---one behind the fixed cylinder and two behind two free-to-move cylinders.

\subsection{V-shaped configuration --- Case (d)}

Another question that arises is how the system behaves if more than one cylinder is placed at any given horizontal location in the wake of the fixed cylinder. In other words, how will the free-to-move cylinders move if we do not remove every other cylinder on each side as we did before? This arrangement is shown in Figure~\ref{fig:V-shape_configurations}(d). The initial distances between neighboring cylinders are $L/D=2$ and $H/D=2$. It is observed that all cylinders initially start moving inward. Since two cylinders are located at each horizontal location, naturally not both of the cylinders can settle right behind the fixed cylinder. The last cylinders on each side change their trajectories from inward to outward at the same time. The third cylinder on the lower side also changes its trajectory later on and moves outward. All the other cylinders keep moving inward to reach an area behind the fixed cylinder, and settle at small vertical distances from the fixed cylinder---as two cylinders exist at each horizontal location and the ones that move toward the centerline must co-exist close to the centerline, and cannot reach the centerline completely. The end configuration consists of a cluster of cylinders right behind the rigid cylinder and three single individual cylinders that are located farther from the centerline from their original locations. With more cylinders on each side of the fixed cylinder, one could imagine that these single cylinders would have been lead cylinders for rows of cylinders that would settle behind them---similar to the cases observed in other configurations of the V-formation here.

\subsection{Flow behavior around the V-shaped configurations}

\begin{figure}
    \centering
    \includegraphics[width=\textwidth]
    {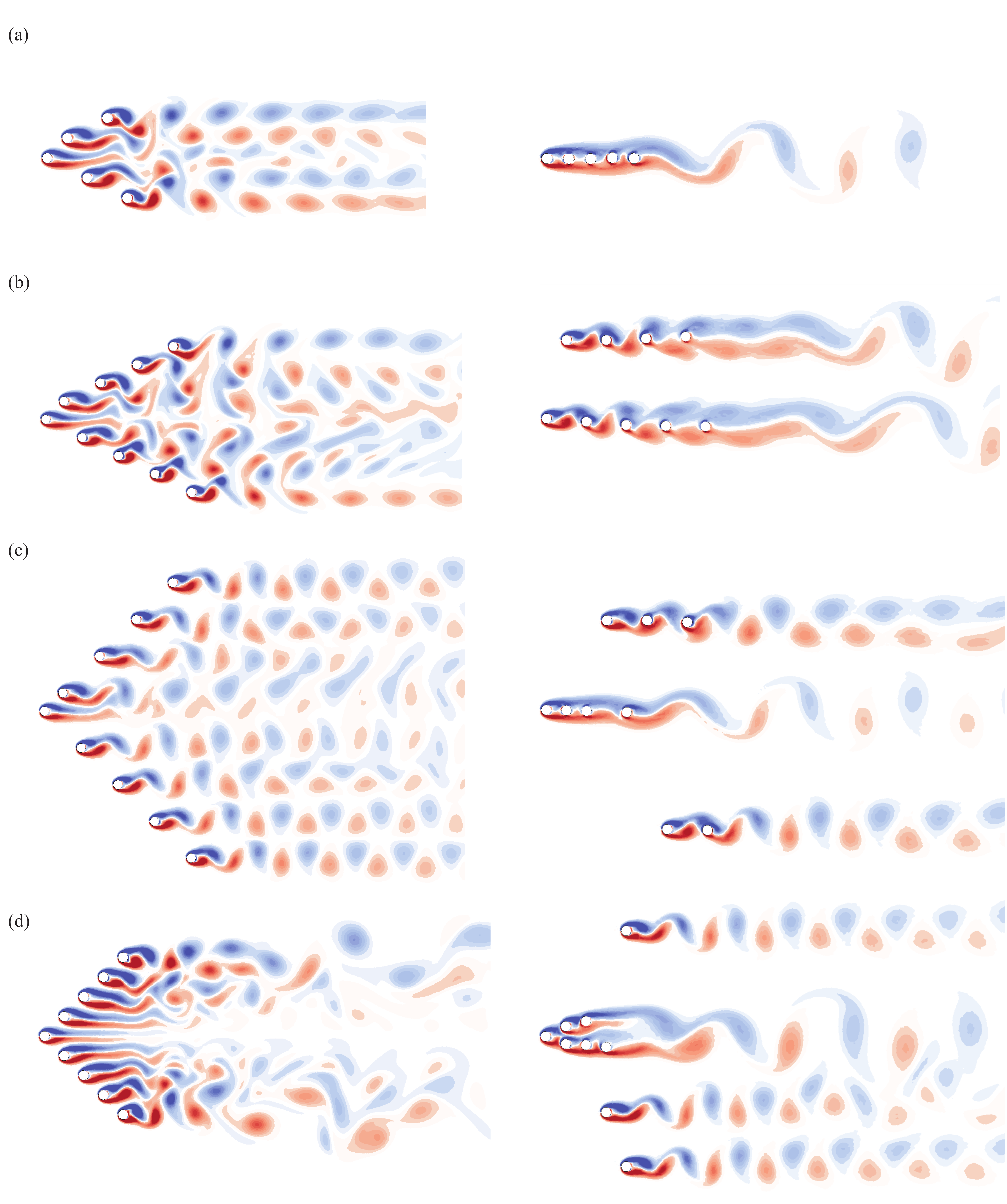}
    \caption{Vorticity contours for the four different V-shaped configurations that are considered here. The left column corresponds to the cases where all cylinders are fixed and the right column corresponds to the steady-state responses of cases where the cylinders in the wake of the first rigid cylinder are free to move in the vertical direction.}\label{fig:V-shape_configurations_fixed_moving}
\end{figure}

The flow behavior around the four V-shaped configurations is shown in Figure~\ref{fig:V-shape_configurations_fixed_moving}. In fixed cases, depending on the proximity of the cylinders, the vortices that are shed from individual cylinders interact with each other. In case (a) two rows of vortex shedding (with a CCW and a CW vortex in each period) are observed on the two sides of the wake together with weak vortices in between. The proximity of the cylinders on the two sides of the configuration has resulted in the formation of one vortex row behind a pair of side cylinders as if the two cylinders act as one single bluff body. In case (b), single vortices (CW in the top row and CCW in the bottom row) are observed at the extremities of the wake, and the vortices that are shed from other cylinders have merged into relatively less organized vortices in between the two single-vortex rows. In case (c) where the distances between cylinders are increased in comparison with the previous cases, two rows of vortex pairs are shed on the two sides of the wake and the vortices that are shed from the five middle cylinders interact with each other. In case (d), where the cylinders are very close to each other, the wake is not organized and no clear pattern is observed in the shed vortices.

After the transient, in cases (a), (b), and (c), vortices are shed behind the one (case (a)), two (case (b)), or three (case (c)) linear configurations that the cylinders have formed. The rows of vortices that are shed from the parallel linear configurations in these cases are far from each other and do not interact---a distinct difference from the case of fixed cylinders. The frequency of shedding behind different rows of cylinders and the size of these vortices depend on the number of cylinders that exist in each row. For example, the vortices of the middle row in case (c) are shed at a lower frequency than the vortices shed behind the two lines of cylinders on the two sides. In case (d), there are three single cylinders and a cluster of cylinders in the middle. The resulting wake consists of four rows of vortices: three of them at the same frequency and with vortices of similar sizes behind the single cylinders, and one at a lower frequency and behind the cluster of cylinders.

\subsection{The influence of the initial conditions on the final configuration}

\begin{figure}
    \centering
    \includegraphics[width=1\textwidth]
    {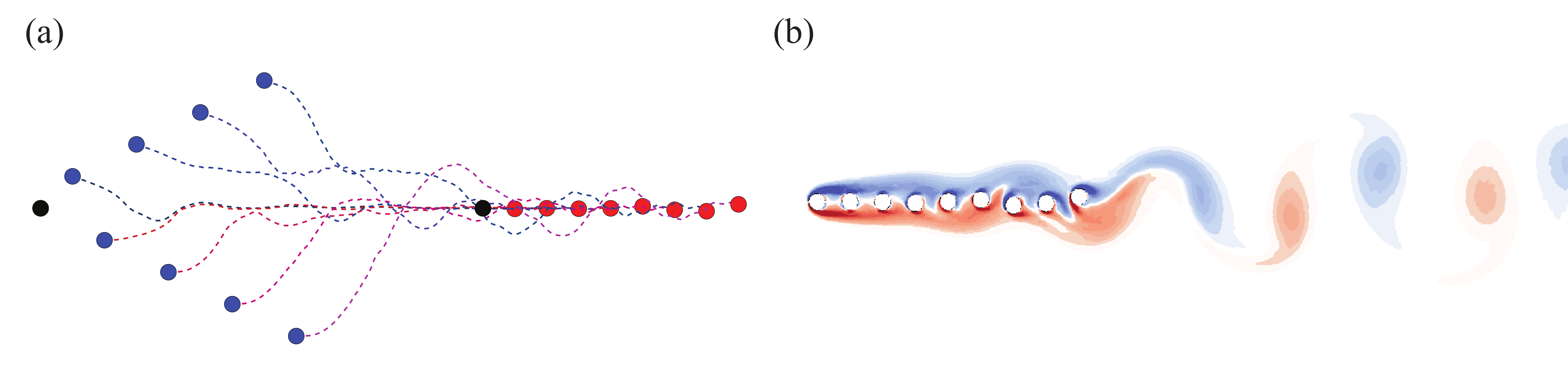}
    \caption{(a) The initial and final locations of cylinders in a V-shaped configuration when a downward initial velocity of $\dot{y} = 0.4 D$/s is imposed on the cylinders on the upper side of the configuration, together with (b) the vorticity contour around the cylinders in their final locations.}
    \label{fig:V-shape_configuration_initialCondition}
\end{figure}

For the V-shaped configurations, the initial conditions also influence the number of linear formations that are formed at the end of the cylinders' transient motion. For example, in the V-shape configuration (b) in Figure~\ref{fig:V-shape_configurations}, these initial conditions cause the initial outward movement of the cylinders on the upper side, which then results in the formation of the second line of cylinders in the final configuration of cylinders. Had this initial condition been influenced such that it would have given an initial inward movement to those cylinders, we predict that all cylinders would have formed one line behind the fixed cylinder. To test this hypothesis in the V-shape configuration (b), we consider a situation where we externally impose such initial conditions on the free-to-move cylinders, i.e., a downward initial velocity of $\dot{y} = 0.4D $/s is imposed on the cylinders on the upper portion of the V-shaped. The results of this test case are shown in Figure~\ref{fig:V-shape_configuration_initialCondition}, where all cylinders on the upper side also move toward the center of the wake and form one single line together with the cylinders on the lower side, behind the fixed cylinder. 

\begin{table}
\centering
\begin{tabular}{ccccc}
\toprule
\multicolumn{1}{c}{} & \multicolumn{2}{c}{Fixed} & \multicolumn{2}{c}{Moving} \\
\cmidrule(rl){2-3} \cmidrule(rl){4-5}
\textbf{} & {$C^*_D$} & {$C^*_L$} & {$C^*_D$} & {$C^*_L$} \\
\midrule
Case (a) & 1.48 & 0.15 & 0.70 & 0.23 \\
Case (b) & 1.53 & 0.31 & 0.57 & 0.60 \\
\bottomrule
\end{tabular}
\caption{Normalized mean drag and fluctuating lift coefficients for cases (a) and (b) of cylinders in the V-shaped configuration.}
\label{fig:table_norm_coeff_Vshape}
\end{table} 

\begin{table}
\centering
\begin{tabular}{ccccc}
\toprule
\multicolumn{1}{c}{} &\multicolumn{2}{c}{Case (a)} & \multicolumn{2}{c}{Case (b)} \\
Cylinder Number & {fixed $C_D$} & {moving $C_D$}& {fixed $C_D$} & {moving $C_D$}  \\
\midrule
1 & 1.33 &  1.17 & 1.23 & 1.28 \\
2 & 1.47 & -0.02 & 1.30 & 1.33\\
3 & 1.55 &  0.23 & 1.39 & 0.51 \\
4 & 1.53 &  1.45 & 1.55 & 0.55 \\
5 & 1.55 & 0.60 & 1.62 & 0.45 \\
6 & - & - & 1.68 & 0.45 \\
7 & - & - & 1.66 & 0.21 \\
8 & - & - & 1.78 & 0.24 \\
9 & - & - & 1.79 & 0.17 \\
\bottomrule
\end{tabular}
\caption{Drag coefficients for cylinders in the V-shaped configuration, when all cylinders are fixed and when the cylinders in the wake are free to move.}
\label{fig:table_norm_coeff_VShape_fixed_moving}
\end{table}

\subsection{Reduced drag in inline formations}

Table~\ref{fig:table_norm_coeff_Vshape} shows how the normalized lift and drag coefficients change for the V-shaped configurations (a) and (b) that we have discussed above. A significant drop in the normalized drag force is clear for both cases, where the drag coefficient decreases from 1.48 to 0.70 for case (a), and from 1.53 to 0.57 for case (b). The drag is reduced to 47\% and 37\% of its original value after the cylinders have been reconfigured to one (for case (a)) or two (for case (b)) lines, respectively. The drag coefficients for each cylinder are given for the cases of fixed and moving V-shaped configurations in Table~\ref{fig:table_norm_coeff_VShape_fixed_moving}. Cylinders that fall in line behind a lead cylinder experience a significant drop in the coefficient of drag. The lead cylinder may also experience a decrease in drag coefficient as in case (a) or experience about the same value as in case (b). This decrease in drag is because the cylinders in their inline formation are close enough to each other that the shear layers that leave the upstream cylinder do not have enough space to form a vortex and therefore travel on the two sides of the lines of cylinders and finally form a vortex downstream the last cylinder of each row. Note that in case (b) cylinders 1 and 2 are the two lead cylinders in the two rows of cylinders and they experience a similar value (and relatively large) drag coefficients. The cylinders in their wakes experience increasingly lower drag coefficients as their locations are farther from the lead cylinders. In case (b), the cylinders are located farther from each other in their state-state condition compared with case (a), and as a result, vortices are formed after the first two cylinders in each row. For the remainder of each row, however, a scenario similar to case (a) is observed where shear layers go around the downstream cylinders and as a result a further reduction in drag is observed for these last cylinders. In all these cases, odd numbers refer to the cylinders behind the fixed cylinder.

\section{Triangular Configuration }

\begin{figure}
    \centering
    \includegraphics[width=\textwidth]
    {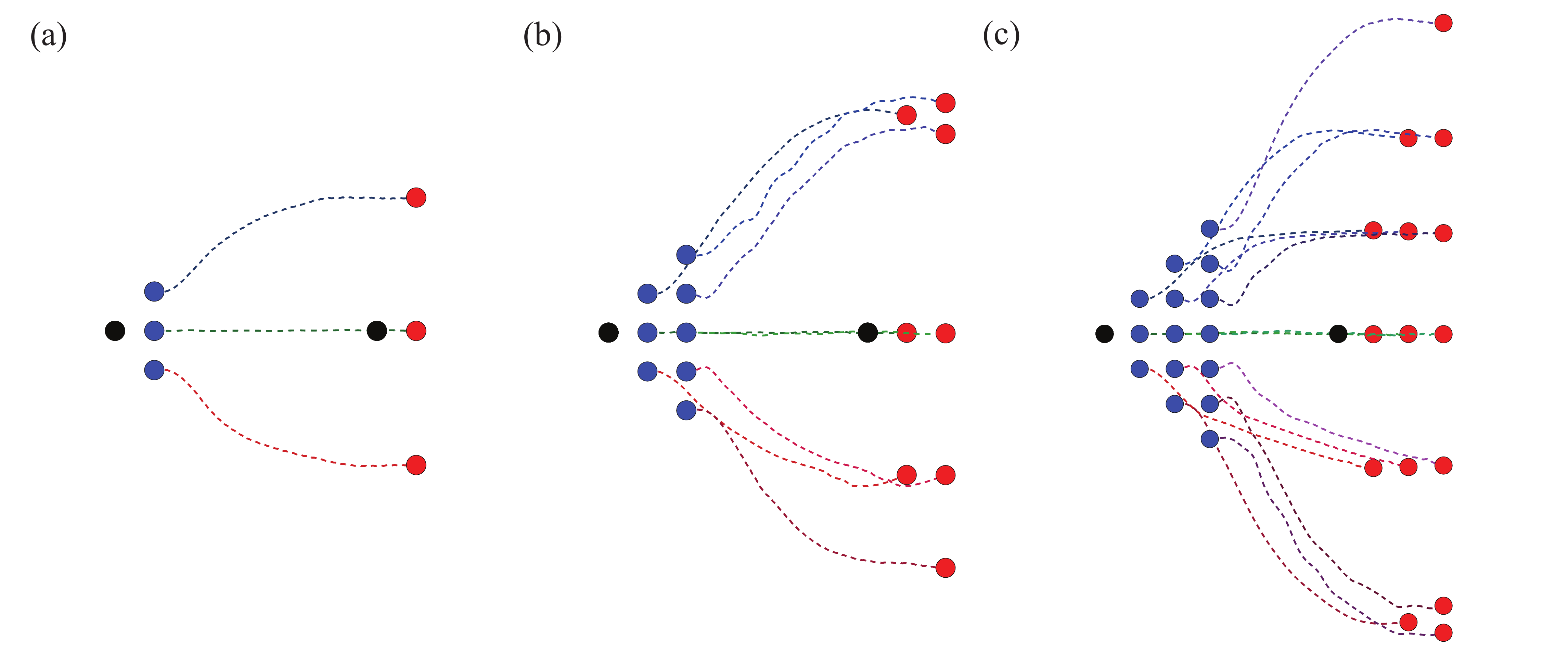}
    \caption{The initial (blue) and the final (red) locations of three cases of ``triangular configurations'' considered here. The dashed lines show the paths that the cylinders take from their initial locations to their final locations.}
    \label{fig:triangular_configurations}
\end{figure}

In the previous two configurations, the rectangular and the V-shaped, the free-to-move cylinders could move inward and form a line behind the fixed cylinder at least as one of their stable steady-state configurations. This was possible mainly because the vertical distance from the original locations of these cylinders and the centerline was not occupied by any other object. In this section and the following section, we change the initial configurations to denser cases in a way that at each horizontal location in the wake of the fixed cylinder, more than two cylinders exist. By doing so, we will observe the behavior of free-to-move cylinders in cases where their possible inward motion is restricted by the presence of other free-to-move cylinders on their ways. We consider two general configurations for these cases: a ``triangular" configuration (in this section) and a ``circular" configuration (in the next section)

\subsection{The overall transient behavior}

We consider three ``triangular'' configurations as shown in Figure~\ref{fig:triangular_configurations} with one, two, and three vertical lines of cylinders located behind the fixed cylinder, resulting in a total of 4, 9, and 16 cylinders, respectively. The distances between all neighboring cylinders are $L/D=2$ and $H/D=2$ for all these cases. This initial configuration results in horizontal lines of cylinders within the overall triangular shape of the configuration. One could expect that if a linear configuration is a stable final configuration for free-to-move cylinders, as we observed in previous cases, these cylinders in horizontal lines will stay together as they transition to their steady-state locations. And since there are several cylinders within the cluster of cylinders, the ones on the upper and the lower sides of the cluster will move outward. This is mainly what is observed in the response of all three of these configurations in Figure~\ref{fig:triangular_configurations} The majority of the horizontal lines of cylinders stay in their horizontal configurations as they move to their final steady-state locations. The exception to this rule is observed in the cases of 9 and 16 cylinders (cases (b) and (c) in the figure), where the single cylinder at the outermost line of the initial configuration merges with the two cylinders in its neighboring line to form a triangular configuration. Similar to the responses of the rectangular and V-shaped configurations, the steady-state configuration of the cylinders in these cases is not symmetric, since the initial condition experienced by the cylinders is not symmetric. In cases where a steady-state triangular configuration is observed on one side, on the other side of the cluster the cylinders stay in two separate lines and do not merge to make a triangular configuration. The clustering behavior seen at the edges of the steady-state configurations of cylinders depends on the initial conditions, and they are just as likely to occur on the top portion of the triangular configuration as on the bottom portion.

\subsection{The flow behavior}

\begin{figure}
    \centering
    \includegraphics[width=\textwidth]
    {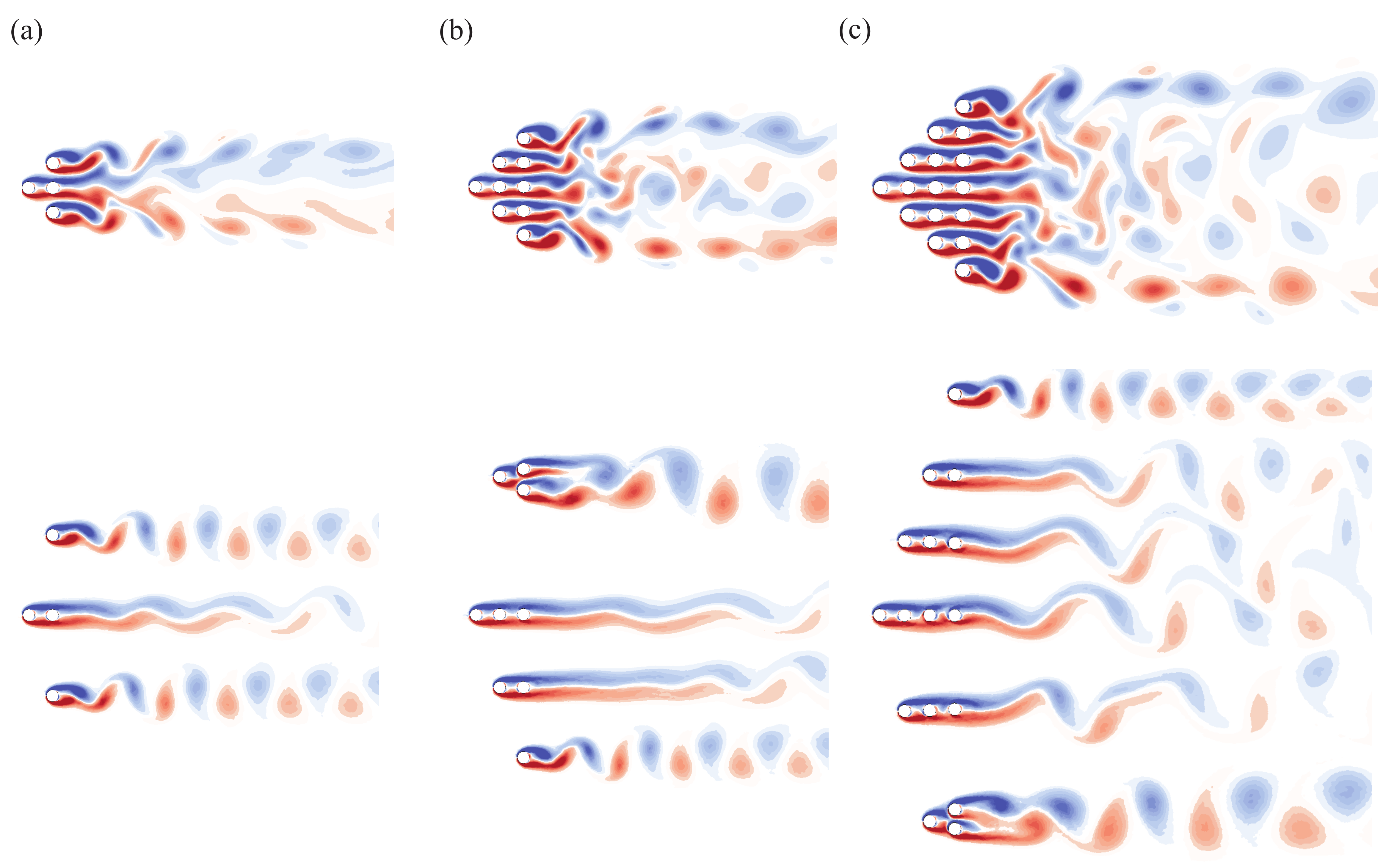}
    \caption{Vorticity contours for the triangular configurations shown in Figure~\ref{fig:triangular_configurations}, for fixed cylinders (upper row) and the final locations of the free-to-move cylinders (lower row).}
    \label{fig:triangular_configurations_vorticity}
\end{figure}

The proximity of cylinders in fixed triangular configurations results in unorganized wakes in all three configurations, as shown in the plots of the first row in Figure~\ref{fig:triangular_configurations_vorticity}. When cylinders are free to move and after they are settled in their inline or small triangular configurations at the end of their transient motion, the wakes resemble those observed in the V-shaped configurations with organized vortex shedding in the wake of each line or cluster. The shedding frequency of these vortices depends on the number of cylinders that have formed each line, and in the case of the cluster cylinders, the number of cylinders that have formed the cluster. In the far wake, several (3, 4, and 6, respectively) rows of cylinders are observed, with vortices shed at different frequencies.

\begin{table}
\centering
\begin{tabular}{ccccc}
\toprule
\multicolumn{1}{c}{} & \multicolumn{2}{c}{Fixed} & \multicolumn{2}{c}{Moving} \\
\cmidrule(rl){2-3} \cmidrule(rl){4-5}
Number of Cylinders & {$C^*_D$} & {$C^*_L$} & {$C^*_D$} & {$C^*_L$} \\
\midrule
4 & 1.33 & 0.01 & 1.14 & 0.14 \\
9 & 1.20 & 0.09 & 0.95 & 0.14 \\
16 & 1.10 & 0.13 & 0.80 & 0.20 \\
\bottomrule
\end{tabular}
\caption{Normalized mean drag and fluctuating lift coefficients for cylinders in triangular configurations.}
\label{fig:table_norm_coeff_triangular}
\end{table}

\begin{figure}
    \centering
    \includegraphics[width=\textwidth]
    {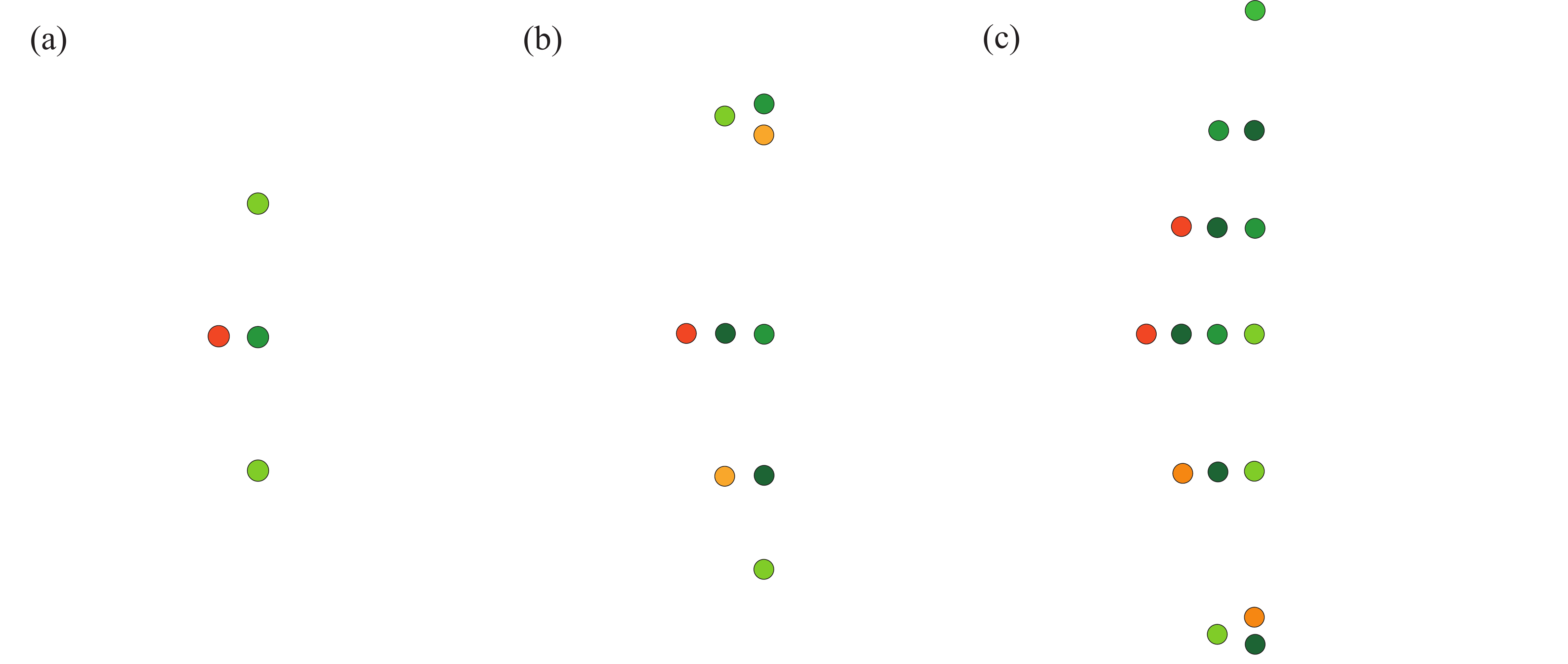}
    \caption{The changes of drag coefficients between fixed and free-to-move configurations, with green indicating a decrease in the drag coefficient and orange indicating an increase. A darker color marks a larger change.}
    \label{fig:circular_coeff_Drag_Lift}
\end{figure}

The reconfiguration of cylinders with the initial triangular configuration also results in a reduction in the overall drag as seen in Table~\ref{fig:table_norm_coeff_triangular}. Here we observe a drag reduction of 86\%, 80\%, and 73\%, respectively for the triangular cases with 4, 9, and 16 cylinders, when they assume their final steady-state configuration in comparison with their original triangular configuration. The drag variations of individual cylinders are given in Table~\ref{fig:table_norm_coeff_triangular_fixed_moving} and shown in Figure~\ref{fig:circular_coeff_Drag_Lift}, where green represents a reduction in drag in a cylinder in comparison with the drag that the same cylinder experienced in the fixed configuration and orange represents an increase in drag. From this plot, it is clear that the cylinders that are located in the wake of other cylinders in an inline configuration experience a reduction in drag. Note that these cylinders were located inline when they were fixed as well, and the reduction in drag is due to the wider distances between rows of cylinders in the final configurations of the free-to-oscillate cylinders, and not solely because they were placed in the wake of a lead cylinder.

\begin{table}
\centering
\begin{tabular}{ccccccc}
\toprule
\multicolumn{1}{c}{} &\multicolumn{2}{c}{4 Cylinders} & \multicolumn{2}{c}{9 Cylinders} &\multicolumn{2}{c}{16 Cylinders} \\
Number of Cylinders & {fixed $C_D$} & {moving $C_D$}& {fixed $C_D$} & {moving $C_D$} & {fixed $C_D$} & {moving $C_D$}  \\
\midrule
1 & 1.05 &  1.34 & 0.94 & 1.33 & 0.87 &  1.34 \\
2 & 1.70 &  1.59 & 1.41 & 1.10 & 1.27 &  1.44 \\
3 & 0.88 &  0.03 & 0.80 & 0.02 & 0.71 &  0.02 \\
4 & 1.70 &  1.60 & 1.42 & 1.43 & 1.29 &  1.42 \\
5 & - & -        & 1.94 & 1.29 & 1.58 &  1.51 \\
6 & - & -        & 0.78 & 1.31 & 0.60 &  0.04 \\
7 & - & -        & 0.78 & 0.32 & 0.85 &  0.31 \\
8 & - & -        & 0.79 & 0.04 & 0.61 &  0.02 \\
9 & - & -        & 1.95 & 1.71 & 1.62 &  1.12 \\
10 & - &  - & - &  -           & 2.09 &  1.80 \\
11 & - &  - & - &  -           & 0.76 &  0.01 \\
12 & - &  - & - &  -           & 0.91 &  0.35 \\
13 & - &  - & - &  -           & 0.63 &  0.41 \\
14 & - &  - & - &  -           & 0.90 &  0.36 \\
15 & - &  - & - &  -           & 0.70 &  1.31 \\
16 & - &  - & - &  -           & 2.14 &  1.36 \\
\bottomrule
\end{tabular}
\caption{Drag coefficients for fixed and free-to-move cylinders in triangular configurations.}
\label{fig:table_norm_coeff_triangular_fixed_moving}
\end{table}

\subsection{The influence of the walls}

\begin{figure}
    \centering
    \includegraphics[width=\textwidth]
    {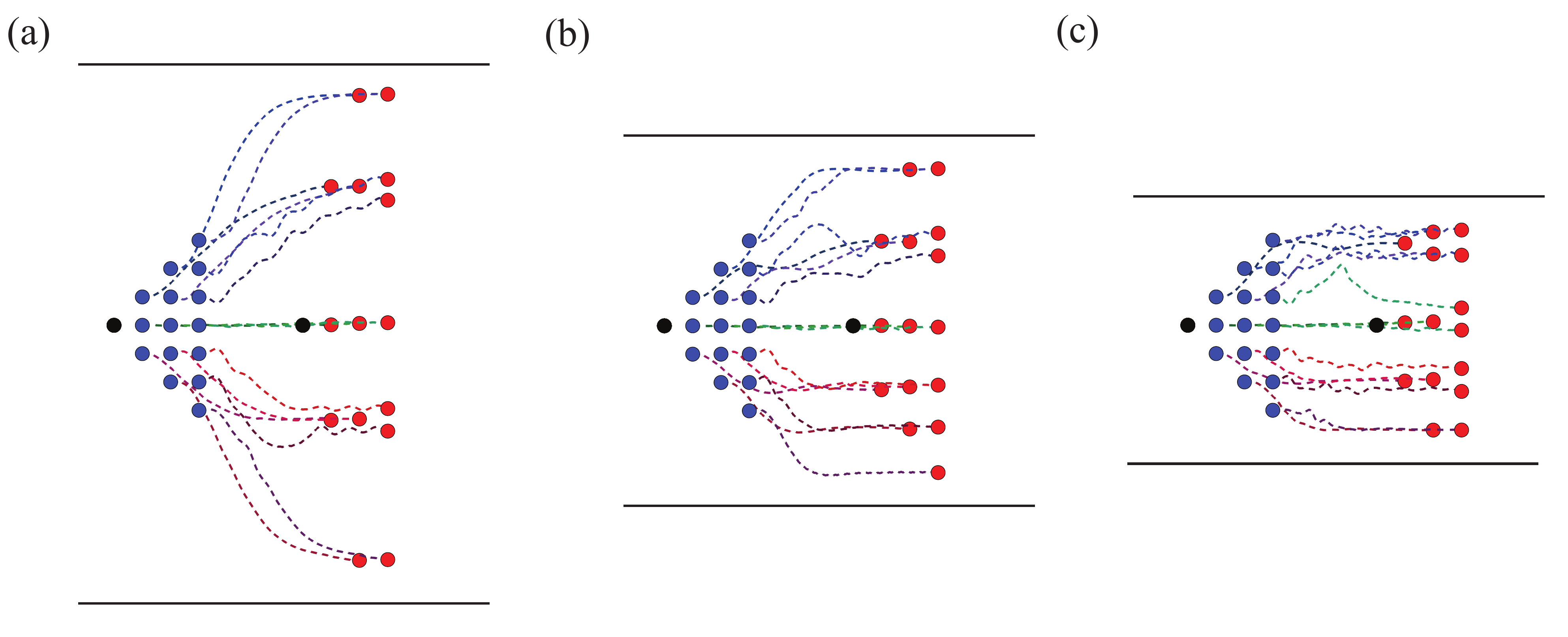}
    \caption{The initial and the final positions of the cylinders in the triangular configuration with sixteen cylinders for blockage ratios of (a) 33\%, (b) 50\%, and (a) 75\%.}
    \label{fig:triangular_blockageratio}
\end{figure}

\begin{figure}
    \centering
    \includegraphics[width=\textwidth]
    {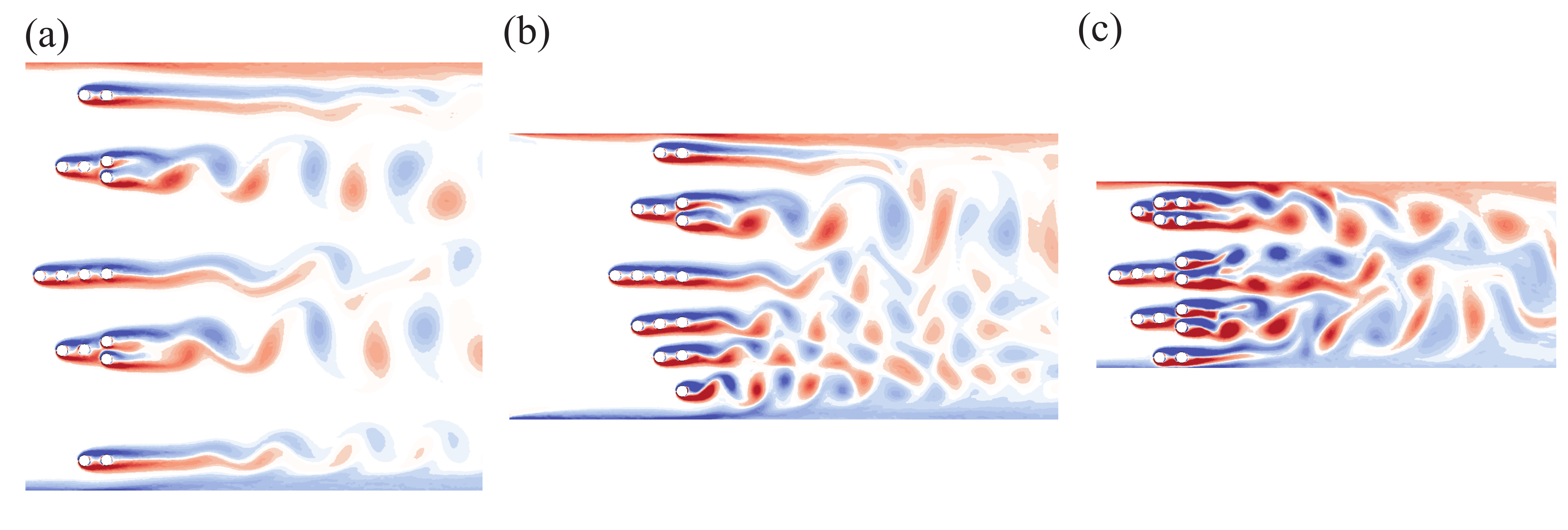}
    \caption{Vorticity contours for the triangular configurations placed within walls as shown in Figure~\ref{fig:triangular_blockageratio}, at the final locations of the free-to-move cylinders.}
    \label{fig:triangular_blockageratio_vorticity}
\end{figure}

In the triangular configurations, the cylinders move outward and reach their steady-state configuration far from the center of the wake. This behavior then begs the question of how the boundaries of the flow domain, or in a physical sense the walls around the configuration, will affect their behavior. How will the presence of the walls and their proximity to the initial configurations of the cylinders will influence their final steady-state configuration? Here, we consider the triangular case with a total of 16 cylinders, and we consider its response for three different wall distances, with blockage ratios of 33\%, 50\%, and 75\%, as shown in Figure~\ref{fig:triangular_blockageratio}. We define the blockage ratio as the distance from the top to the bottom cylinder in the triangular configuration over the height of the domain. In all three cases, the cylinders still try to form a linear configuration, albeit in a smaller and smaller region, as the blockage ratio is increased. The lines of cylinders get closer to each other with an increased blockage ratio. In all cases, besides the single lines, lines with triangular endings form as well. In the case with the maximum blockage ratio, three different steady-state configurations are observed: the linear configuration, the linear configuration with the triangular ending, and the rectangular configuration. An increased blockage ratio results in increased overall drag force on the cluster of cylinders (Table~\ref{fig:table_norm_coeff_triangular_blockage}). This is similar to the case of a single cylinder in channel flow, in which as the blockage ratio is increased, the drag coefficient increases~\citep{Norberg1994, Anagnostopoulos1996, Anagnostopoulos2004, Chakraborty2004, Sahin2004, Kumar2006}. For the smallest blockage ratio, the rows of cylinders move away from each other significantly---similar to what they did in the configuration with no wall. As a result, the normalized drag is very similar to what we observed in the original case, and the wall effects are minimal. For the other two blockage ratios, the normalized drag increases with increasing blockage ratio. This can be explained by the fact that the rows of cylinders cannot move away from each other in these higher blockage ratio cases, and as a result, the large normalized drag that they experienced in their fixed cases is not reduced.

The wake patterns observed in these cases are shown in Figure~\ref{fig:triangular_blockageratio_vorticity}. In the case with the smallest blockage ratio, case (a),  the wake is very similar to what we observed in the case with no walls: the rows of cylinders are relatively far from each other, and relatively organized vortices are formed in the wakes of the rows of cylinders and small clusters of cylinders. As the blockage ratio is increased, the organized shedding is less and less the case, and the vortices start interacting with each other in the wake. The case with the maximum blockage ratio exhibits an unorganized wake, very similar to what we observed in the fixed case of this configuration, since with this high blockage ratio, the cylinders stay close to each other and resemble their original configuration even after the transient.

\begin{table}
\centering
\begin{tabular}{ccccc}
\toprule
\multicolumn{1}{c}{} & \multicolumn{4}{c}{16 Cylinder Case} \\
\cmidrule(rl){2-5} 
Blockage ratio & \multicolumn{2}{c}{$C^*_D$} &\multicolumn{2}{c} {$C^*_L$} \\
\midrule
33\% & \multicolumn{2}{c} {0.778} &  \multicolumn{2}{c} {0.091}  \\
50\% & \multicolumn{2}{c} {1.086} &  \multicolumn{2}{c} {0.128} \\
75\% & \multicolumn{2}{c} {1.861} &  \multicolumn{2}{c} {0.132}  \\
\bottomrule
\end{tabular}
\caption{Normalized mean drag and fluctuating lift coefficients for cylinders in triangular configurations under three blockage ratios.}
\label{fig:table_norm_coeff_triangular_blockage}
\end{table}

\section{Circular Configurations}

\begin{figure}
    \centering
    \includegraphics[width=\textwidth]
    {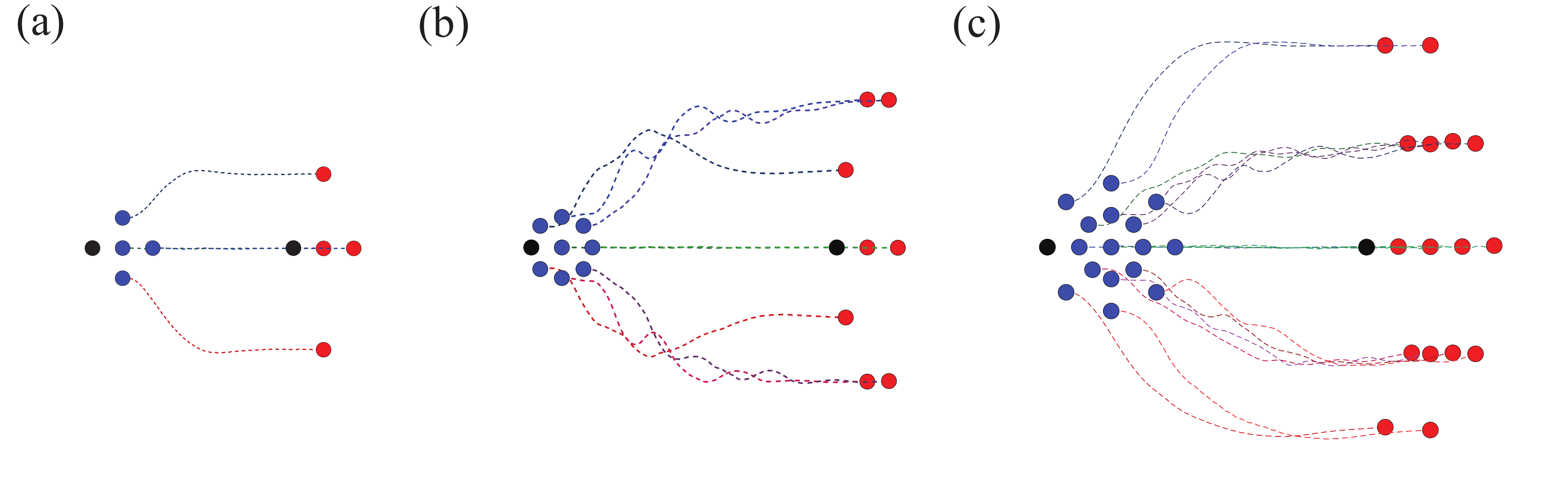}
    \caption{The initial (blue) and the final (red) locations of four cases of ``circular configurations'' considered here. The dashed lines show the paths that the cylinders take from their initial locations to their final locations.}
    \label{fig:circular_configurations}
\end{figure}

In this section, we consider another ``compact'' configuration for a cluster of cylinders in flow: a circular configuration, in which the cylinders are originally located at equal radial distances from the cylinder at the center. The distances between two immediate cylinders in this configuration are $R/D=2$, where $R$ represents the distance in the radial direction. We consider three such configurations here, as shown in Figure~\ref{fig:circular_configurations}, with an increasing number of cylinders in the cluster. In case (a), the middle row of cylinders stays at its original location, as expected, and the two cylinders on the sides move to a steady-state location farther from the centerline. This is much expected after observing the behavior of free-to-move cylinders in the triangular configuration.
In cases (b) and (c), one can observe ``curved'' rows of cylinders, similar to the rows that formed the triangular configurations in the previous section. It is interesting to observe that these initial curved rows do not stay together during the transient. In case (b), the leading cylinder on each side transitions to a solitary final position, while the two cylinders that were originally placed in the wake of this cylinder form a line in the final configuration. The cylinder that was originally placed partially behind the lead cylinder takes the lead in the two-cylinder line. In case (d), it is predictable that the top two cylinders could form their line since they are placed relatively far from the rest of the group. The other four free-to-move cylinders on each side then form their line. Although there are differences in the details of how these cylinders form lines, overall, they behave very similarly to the triangular case: they move outward due to the high density of the cylinder population centrally in the configuration and form lines where the mean lift force becomes negligible. Similar to the triangular configurations, the overall drag force that acts on the cylinders decreases for the final linear configurations of the cylinders in comparison with their original locations as evidenced by the numbers given in Table~\ref{fig:table_norm_coeff_circular}.

\begin{figure}
    \centering
    \includegraphics[width=\textwidth]
    {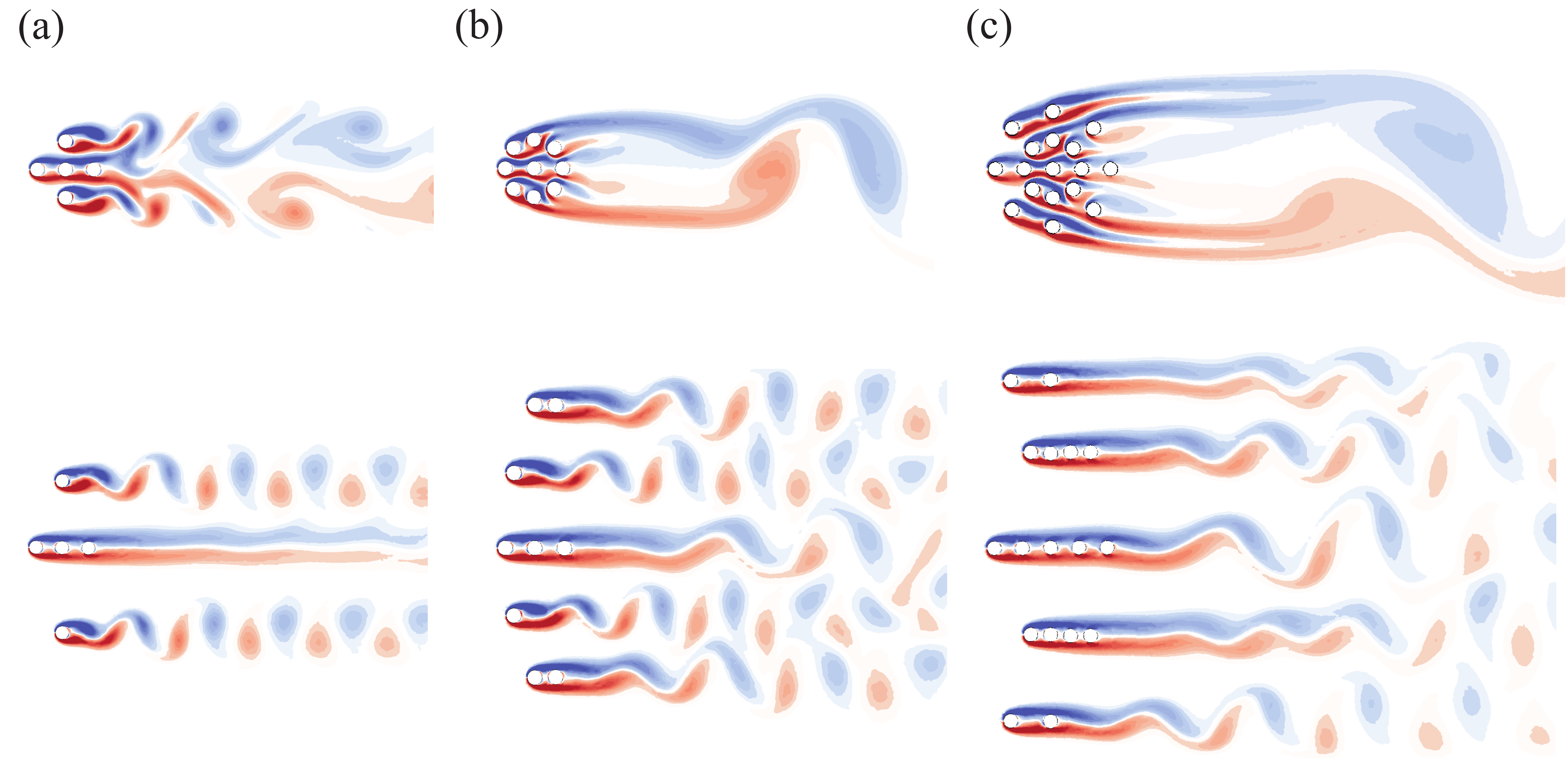}
    \caption{Vorticity contours for the circular configurations shown in Figure~\ref{fig:circular_configurations}, for fixed configurations (upper row) and the final locations of the moving configurations (lower row).}
    \label{fig:circular_configurations_vorticity}
\end{figure}

The wakes of these cluster cylinders for when they are fixed are shown in the top row of Figure~\ref{fig:circular_configurations_vorticity}. In case (a) where there are only 4 cylinders in the cluster, the wake is similar to the wake we observed in the triangular configurations: vortices that are shed from each row of cylinders interact with each other and form a relatively unorganized wake. For cases (b) and (c), the high density of cylinders results in the cluster acting as a bluff body, and large vortices are observed in the wake. This is in agreement with previous observations by \cite{Hosseini2020} where they also observed large-scale shedding in the wake of such clusters of fixed cylinders. When the cylinders are free-to-move, however, they form lines with different numbers of cylinders, very similar to what they did in the case of triangular configurations, and therefore it is not surprising to observe several (3, 5, and 5, respectively) rows of vortices in the wake of these cylinders when they reach steady state---again similar to what we had observed for the triangular configurations. Note that these linear steady-state configurations are achieved despite very different initial wake patterns---unorganized wakes in case (a), versus organized large-scale vortices in cases (b) and (c).

\begin{table}
\centering
\begin{tabular}{ccccc}
\toprule
\multicolumn{1}{c}{} & \multicolumn{2}{c}{Fixed} & \multicolumn{2}{c}{Moving} \\
\cmidrule(rl){2-3} \cmidrule(rl){4-5}
Number of Cylinders & {$C^*_D$} & {$C^*_L$} & {$C^*_D$} & {$C^*_L$} \\
\midrule
5 & 1.09 & 0.05 & 0.95 & 0.10 \\
9 & 0.81 & 0.04 & 0.59 & 0.20 \\
17 & 0.83 & 0.001 & 0.56 & 0.091 \\
\bottomrule
\end{tabular}
\caption{Normalized mean drag and fluctuating lift coefficients for cylinders in circular configurations.}
\label{fig:table_norm_coeff_circular}
\end{table}

\section{The influence of the mass ratio}

We chose a relatively large mass ratio, $m^*=12.7$, for the cases that we have discussed so far so that the main motion that is observed would be that of the cylinders' transient from their original locations to their final locations, with minimal oscillations. Here, we show how a low mass ratio case, i.e. $m^*=1$, behaves when the cylinders are placed in the same configurations as some of the cases we discussed before.

\begin{figure}
    \centering
    \includegraphics[width=0.7\textwidth]
    {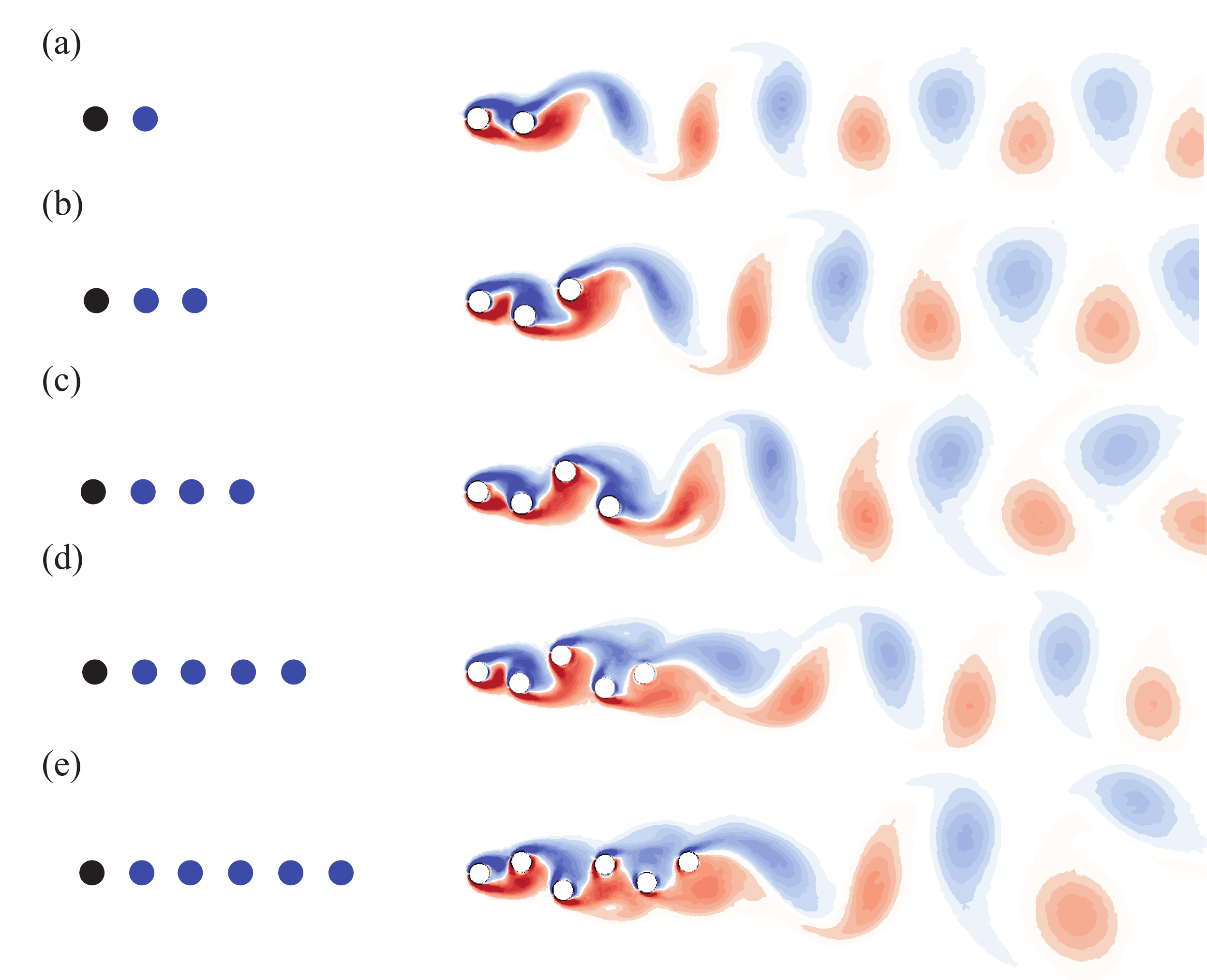}
    \caption{Vorticity plots for inline configurations with a mass ratio of 1, for (a) 2, (b) 3, (c) 4, (d) 5, and (d) 6 cylinders.}
    \label{fig:inline_configurations_vorticity_mass_ratio_1}
\end{figure}

We first consider the inline configurations, as we have observed from our high mass ratio results that the inline configuration is the desired final steady-state configuration for many initial configurations of the cluster of cylinders. The question is whether we observe a major difference in the response of the inline configuration if we decrease the mass ratio by more than an order of magnitude. The results are shown in Figure ~\ref{fig:inline_configurations_vorticity_mass_ratio_1}. This figure is equivalent to Figure~\ref{fig:Inline_configurations}, but for a mass ratio of 1. As observed in the figure, the smaller mass ratio results in oscillations of cylinders with relatively large amplitudes of oscillations (approximately $2D$). These large-amplitude oscillations result in the shedding of vortices in the wake closer to the cylinders for these cases in comparison with the large mass ratio cases of Figure~\ref{fig:Inline_configurations} where the vortices were shed farther downstream. Despite their large-amplitude oscillations, the cylinder's mean displacements stay close to zero in the low-mass ratio cases, implying that the cylinders, on average, stay in the wake of the fixed cylinder, as they did in the large-mass ratio cases. The only difference in the motion of these cylinders, therefore, is their large-amplitude oscillations---they still stay behind the fixed cylinder.

\begin{figure}
    \centering
    \includegraphics[width=\textwidth]
    {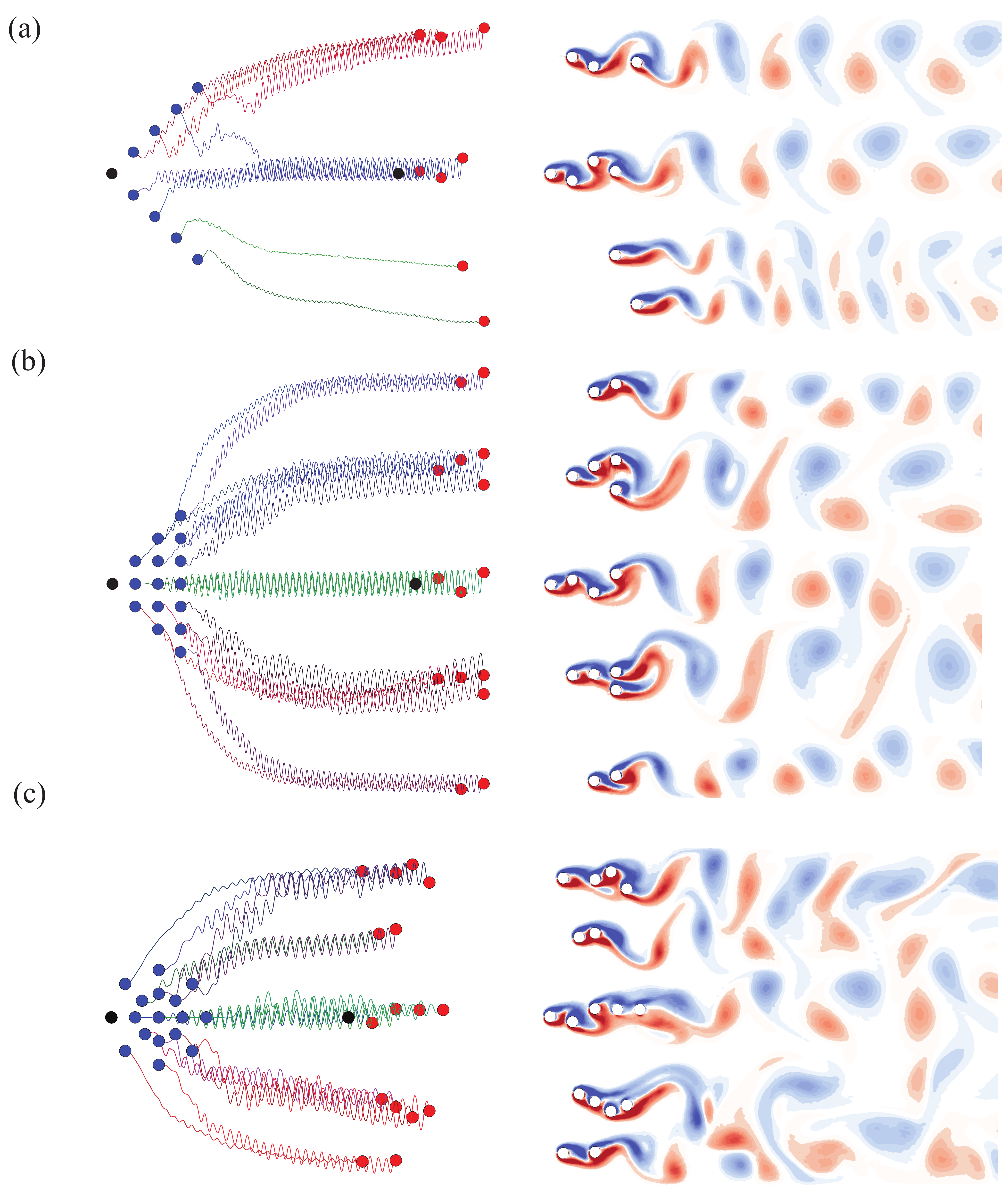}
    \caption{The transient behavior of the (a) V-shaped, (b) triangular, and (c) circular configurations with a mass ratio of 1. Transitions from the initial position to the final position are shown on the left and snapshots of vorticity contours in steady state on the right.}
    \label{fig:self-assemby_massRatio1}
\end{figure}

The question then arises whether starting from the other configurations (V-shaped, triangular, and circular) will influence the transient and the final steady-state locations of the low-mass-ratio cylinders. To answer this question, we consider the cases with the highest number of cylinders from the V-shaped, triangular, and circular configurations. The cylinders' behavior and their wake patterns are shown in Figure~\ref{fig:self-assemby_massRatio1}. In these cases, the paths taken by the cylinders toward their final steady-state position are a superposition of a mean displacement (similar to what they did in the large-mass-ratio cases) and a periodic response about this mean displacement. The amplitude of these oscillations could be very small, such as those observed for the two lower cylinders in the V-shaped case, but in general, these oscillations reach amplitudes of approximately $2D$. The mean path that each cylinder takes, however, is very similar to the mean path that a cylinder with a higher mass ratio would have taken. Again the dependence on the initial conditions would result in slight variations in the final locations of cylinders, however, in general, the paths are very similar in these configurations for the low- and high-mass ratio cases. The wake, on the other hand, looks very different. Due to the large-amplitude oscillations of cylinders, the vortices observed in the wake of each row of cylinders in their final positions interact with each other more intensely than they did in the large-mass-ratio case. These large-amplitude oscillations also result in some differences in the final configurations of cylinders in terms of the number of cylinders observed in each row. For example, in the circular configuration at the low mass ratio, the upper row consists of four cylinders, instead of three which we had observed in the high-mass-ratio case.

Note that the only cylinders that oscillate with small amplitudes in their final configurations are the two cylinders in the lower rows of the V-shaped. Each of these cylinders remains as a single cylinder in each row, and during their entire paths toward these final positions, they have stayed far from other cylinders. This then implies that oscillations with large amplitude are caused due to the interactions of the wakes of cylinders placed in a row, which in turn results in the shedding of larger vortices in their wakes and therefore large-amplitude oscillations. This can also be observed in the time histories of other cylinders: Toward the beginning of their motions, the cylinders that start far from other cylinders oscillate with small amplitudes. As soon as they reach the vicinity of another cylinder, the amplitudes of their oscillations increase, as their wake interacts with the wake of the neighboring cylinder, and form larger vortices.

Overall, while there are differences in the details of the cylinders' motion when the mass ratio is decreased, the overall transient trend that is observed (the average paths that the cylinders take, and the linear configuration that they form at the end of their transient motion) does not change significantly by changing the mass ratio---at least within the range that we have considered here.

\section{Conclusions}

We study the collective behavior of a cluster of cylinders placed in flow and free to move perpendicular to the direction of flow---except for the very first cylinder that remains rigid. The first cylinder disturbs the incoming flow and would have shed vortices downstream if there were no cylinders in its wake. In the cases we consider here, however, cylinders are placed in the wake of this rigid cylinder. No structural stiffness or damping exists in the direction of motion, and the cylinders are not allowed to move in the direction of flow---so that they would not be washed downstream by the incoming flow. We have considered five general configurations: linear, rectangular, V-shaped, triangular, and circular, and for each configuration we have considered different number of cylinders in the cluster. These clusters are inspired by the types of formations that are observed in the collective behavior of animals, birds, and fishes, with a general question of how bluff bodies would have behaved in these collective scenarios.  

We have studied the flow forces that act on these cylinders and the flow behavior around them for the configurations with free-to-move cylinders in the wake of the fixed cylinder as well as for cases where we keep all cylinders fixed in their place to have a basis for comparison. For cases with free-to-move cylinders, we have followed the paths of these cylinders as they move toward their final positions. Overall, and independent of the original configuration, a linear (or very close to linear) configuration is the most desired final configuration for all these cases. In configurations where at each horizontal distance from the fixed cylinder only one free-to-move cylinder exists, linear formations are easier to achieve, due to the lack of competition between two cylinders to reach the same spot. In cases where more than one cylinder exists at a given horizontal distance from the fixed cylinder, the competition between free-to-move cylinders sometimes results in formations that include two of these cylinders next to each other. The steady state is achieved for free-to-move cylinders when the mean value of lift that acts on them becomes zero. Clearly, at their original positions in a formation, these mean values of lifts are not necessarily zero. It is this non-zero lift that forces the cylinders to move away from their original locations---toward the centerline of the wake or away from it---and move toward their steady-state position. The steady-state positions are not necessarily symmetric with respect to the rigid cylinder, even for symmetric original configurations. This is because the pressure distribution around the cylinders that are placed in the wake of the rigid cylinder is not symmetric due to the asymmetric shedding from the fixed cylinder. This transient motion can be influenced by influencing the initial conditions that the cylinders see. Forcing them initially to move in a particular direction (by externally applying some initial velocity in a desired direction) creates a competition between these externally imposed ``forces" and the mean lift force that is exerted on the cylinder due to the asymmetric pressure distribution around them as they are placed in the wake of the fixed cylinder. As a byproduct of the transient motion of the free-to-move cylinders and their reconfiguration to linear formations, the total drag per cylinder (a normalized drag coefficient as we have defined in this work) is reduced. The exact amount of drag reduction in different configurations changes, but in all configurations drag reduction does occur. The number of the final linear configurations depends on the original configuration and the number of cylinders. It is not that independent from the configurations all cylinders would from one single line behind the fixed cylinder. In many cases, one of the free-to-move cylinders takes the lead on a new line of cylinders. We have done the major part of this study for cylinders with a relatively large mass ratio. We show that for a low mass ratio, naturally, the cylinders feel ``lighter'' and as a result respond to the vortices that interact with them and oscillate, but still they follow very similar paths to those taken by the cylinders in the high-mass-ratio cases and converge to a series of linear configurations at the end of their transition. 

\bibliographystyle{unsrtnat}

\bibliography{arxiv-paper}

\end{document}